\documentclass[12pt]{article}
\usepackage{latexsym}
\usepackage{epic}
\usepackage{eepic}

\usepackage{backref}

\usepackage{trans000}       
\newcommand{\restrict}{|}

\begin{document}


\title{MSO definable string transductions\\
and
two-way finite state transducers}

\author{Joost Engelfriet \quad and \quad Hendrik Jan Hoogeboom \\
Leiden University, \hspace{1ex}
Institute of Computer Science\\
P.O. Box 9512, 2300 RA Leiden,
The Netherlands
}
\date{\small Technical Report 98-13, December 1998}

\maketitle

\begin{abstract}
String transductions that are definable in monadic second-order (mso) logic
(without the use of parameters) are exactly those realized by
deterministic two-way finite state transducers.
Nondeterministic mso definable string transductions
(i.e., those definable with the use of parameters)
correspond to
compositions of two nondeterministic two-way finite state transducers
that have the finite visit property.
Both families of mso definable string transductions
are characterized in terms of Hennie machines,
i.e., two-way finite state transducers with the finite visit property
that are allowed
to rewrite their input tape. 
\end{abstract}

\section*{Introduction}

In language theory,
it is always a pleasant surprise when two formalisms,
introduced with different motivations,
turn out to be equally powerful,
as this
indicates that the underlying concept is a natural one.
Additionally,
this means that notions and tools from one formalism can be made
use of within the other,
leading to a better understanding of the formalisms under consideration.
Most famous in this respect
are of course the regular languages \cite{yu},
that can be defined
using a computational formalism
(finite state automata, either deterministic or nondeterministic),
but also have well-known
grammatical (right-linear grammars),
operational (rational operations),
algebraic (congruences of finite index),
and logical (monadic second-order logic of one successor)
characterizations
\cite{mcculloch,2to1a,chomsky,kleene,myhill,nerode,buchi,elgot}.

In this paper we study `regular' (string-to-string) transductions,
rather than regular languages,
and we obtain the equivalence of particular computational
and logical formalisms,
modestly following in the footsteps of B\"uchi and Elgot.
Their original work \cite{buchi,elgot},
demonstrating how a logical formula may effectively be transformed
into a finite state automaton accepting the language
specified by the formula when interpreted over finite sequences,
shows  how to relate the {\em specification} of a system behaviour
(as given by the formula)
to a possible {\em implementation}
(as the finite state behaviour of an automaton).
In recent years much effort has been put into
transforming these initial theoretical results
into software tools for the verification of finite state systems,
{\em model checking}, see the monograph \cite{kurshan}.
Generalizations of the result of B\"uchi and Elgot
include
infinite strings \cite{buchi62},
trees \cite{doner,thwr},
traces (a syntactic model for concurrency) \cite{diekroz},
texts (strings with an additional ordering) \cite{htp},
and tree-to-tree transductions \cite{bloem,maneth}.
We refer to \cite{handbook-thomas} for an overview of
the study of formal languages within the framework of mathematical logic.

\medskip

We give a short description of the two formalisms of `regular'
string transductions
that we study in this paper.
We mainly consider the deterministic case.
\medskip


A two-way finite state transducer
(or two-way generalized sequential machine, 2gsm)
is a finite state automaton equipped with a two-way input tape,
and a one-way output tape.
Such a transducer may freely move over its input tape,
and may typically reverse or copy parts of its input string.
It is, e.g., straightforward to construct a transducer
realizing the relation
$\{ (w,ww) \mid w\in \{a,b\}^* \}$.
It should be clear from this example that
regular languages are not closed under 2gsm mappings,
contrary to their closure under one-way gsm mappings.

However, it is well known
\cite{2to1a,2to1b,houl-book}
that two-way finite state automata accept only regular languages,
and consequently (using a straightforward direct product construction)
the regular languages are closed under
inverse 2gsm transductions.
From this general result we may infer a large number of
specific closure properties of the regular languages,
such as closure under the `root' operation
${}^{\sqrt{}}K = \{ w \mid ww \in K \}$.
It is maybe less well known that the (deterministic) 2gsm mappings
are closed under composition \cite{chytil}.
This result is used as a powerful tool in this paper.

\medskip

The monadic second-order (mso) logic of one successor
is a logical framework that allows one to specify
string properties using quantification over
sets of positions in the string.
As stated above, B\"uchi and Elgot proved that the
string languages  specified by mso definable
properties are exactly the regular languages.
The logic has a natural generalization to graphs, with
quantification over sets of nodes,
and predicates referring to node labels
and edge labels.
It is used to define graph-to-graph transductions,
by specifying the edges of the output graph
in terms of properties of (copies of) a given input graph
\cite{cou97,joost-handbook}.
This is just a special case of the notion of interpretation
of logical structures, well known in mathematical logic
(see, e.g., \cite[Section~6]{seese}). These
 mso definable graph transductions play an important role 
 in the theory of graph rewriting,
as the two main families of context-free graph languages
can be obtained by applying mso definable graph transductions
to regular tree languages \cite{engooslog,hypergraphs}.

Here we consider
mso definable string transductions,
i.e., the restriction of mso definable graph transductions to
linear input and output graphs.
It is known that mso definable (string) transductions
are closed under composition, and that the regular languages
are closed under inverse mso definable transductions
(recall that regular is equivalent to mso definable),
see, e.g., \cite{msotrans}.

\medskip
Apart from these similar closure properties there is more
evidence in the literature that indicates the close connection
between 2gsm transductions and mso definable transductions.
First, various specific 2gsm transductions were shown to be
mso definable, such as one-way gsm mappings,
mirror image, and
mapping the string $w$ onto $w^n$ (for fixed $n$),
cf. \cite[Prop~5.5.3]{cou97}.
Second, 
returning to the theory of graph grammars,
it is explained in
\cite[pages 192--8]{joost-handbook} that
the ranges (i.e., output languages)
 of mso definable (string) transductions are equal to the
(string) languages defined by linear context-free graph grammars,
which, by a result of
\cite{heyker},
equal the ranges of 2gsm transductions.
Consequently, the two families of transductions we consider
have the same generative power (on regular input).
This, however, does not answer the question whether they are
the same family of transductions (cf. Section~6 of \cite{msotrans}).
In this paper we answer this question positively
(in the deterministic case).
Thus, string transductions that are {\em specified} in mso logic
can be {\em implemented} on 2gsm's, and vice versa.

\medskip
Our paper is organized as follows.

In a preliminary section we mainly recall notions and notations
regarding graphs, in particular mso logic for graphs and strings.
Moreover, we recall the usual, natural
representation of strings as linear graphs that allows 
a transparent interpretation of strings and string languages within
the setting of the mso logic for graphs.

In Section~\ref{sect:2way} we study {\em two-way machines},
our incarnation of two-way generalized sequential machines.
We extend the basic model by allowing the machines to `jump' to new
positions on the tape (not necessarily adjacent to the present
position) as specified by an mso formula that is part of the
instructions. This `hybrid' model (in between logic and machine)
facilitates the proof of our main result.
We consider yet
another variant of the 2gsm which allows `regular look-around',
i.e., the ability to test the strings to the left and to the right
of the reading head for membership in a regular language.
The equivalence of the basic 2gsm model and our two extended models
(in the deterministic case)
is demonstrated using the closure of  2gsm under composition
and using B\"uchi and Elgot's result for regular languages.

In Section~\ref{sect:mso} we recall the definition of mso definable graph
transduction, and restrict that general
notion to mso definable string transductions by considering 
graph representations for strings.
In addition to the representation of Section~\ref{preliminaries},
we use an alternative,
natural and well-known,  graph representation for strings.
Again it uses linear graphs, with labels 
on the edges rather than on the nodes to represent the
symbols of the string.
These two representations differ slightly,
due to an unfortunate minor technicality involving the empty string;
the second representation gives more uniform results.

The main result of the paper is presented as Theorem~\ref{thm-logic&machine}:
the equivalence of
the (deterministic) 2gsm  from Section~\ref{sect:2way},
and the mso definable string transductions
from Section~\ref{sect:mso}. Section~\ref{sect:logic&machine}
contains the proof of this result.
In order to transform a 2gsm into the mso formalism we consider
the `computation space' of a 2gsm on a given input.
This is the graph which has a node for each pair consisting of a tape position
and a state of the 2gsm. These nodes are connected by edges representing the
possible
moves of the 2gsm. The transduction is then decomposed into (basically) two
constructions, each of which is shown to be mso definable. First the
computation space is defined in terms of the input string, then
the computation path for the input (and its resulting output string)
is recovered from the computation graph. One implication of the main
result then follows by the closure of mso definable (graph!) transductions
under composition.
The reverse implication is obtained by transforming an mso definable string
transduction into a 2gsm equipped with mso instructions, the tool we
introduced in Section~\ref{sect:2way}.

In Section~\ref{sect:nondet}
we study nondeterminism.
This feature can be added to mso definable transductions by introducing
so-called `parameters': free set variables in the definition of the
transduction \cite{cou97}. The output of the transduction for a given input
may then vary for different
valuations of these parameters.
These transductions are closed under composition, as opposed to those
realized by nondeterministic 2gsm. We conclude that
as opposed to the deterministic case,
the two nondeterministic families are incomparable.
Finally, we observe that the family of nondeterministic mso transductions
is equal to the family of transductions defined by composing
a (nondeterministic) relabelling and a deterministic transduction.

Finite visit machines form the topic of our final section,
Section~\ref{finite-visit}.
These machines have a fixed bound on the number of times each of
the positions of their input tape may be visited
during a computation.
We characterize the nondeterministic mso definable string transductions
as compositions of two nondeterministic
2gsm's with the finite visit property.
Additionally we demonstrate that an arbitrary composition of nondeterministic
2gsm's realizes a nondeterministic mso definable string transduction
if and only if that transduction is finitary, i.e.,
it has a finite number of images for every input string.

A more direct characterization can be obtained by considering
Hennie transducers, i.e.,
finite visit
2gsm's that are allowed to rewrite the symbols on their
input tape.
These machines characterize  the mso
definable transductions,
both in the deterministic case \cite{chytil}
and the nondeterministic case.

\bigskip
An extended abstract of this paper is published as
\cite{icalp}.

\section{Preliminaries}\label{preliminaries}

We recall some notions and results
regarding graphs and their monadic second order logic.

By $|w|$ we denote the length of the string $w$.

We use $\cmp$ to denote the composition of binary relations 
(note the order):
$R_1 \cmp R_2 = \{ (w_1, w_3) \mid
\mbox{ there exists   $w_2$ such that } (w_1,w_2) \in R_1,
(w_2,w_3) \in R_2 \}$, and extend it to
families of binary relations:
$F_1 \cmp F_2 = \{ R_1 \cmp R_2 \mid
R_1 \in F_1, R_2 \in F_2 \}$.

A binary relation $R$ is {\em functional},
if $(w,z_1)\in R$ and $(w,z_2)\in R$ imply $z_1 = z_2$.
It is {\em finitary}, if each original is mapped to only
finitely many images, i.e., the set
$\{ z \mid (w,z) \in R \}$ is finite for each $w$ in the domain of $R$.

\paragraph{Graphs.}
Let  $\Sigma$  and  $\Gamma$ be alphabets
of node labels	and edge labels, respectively.
A {\em graph} over $\Sigma$ and $\Gamma$ is a triple $g = (V,E,{\labl})$,
where $V$ is the finite set of nodes,
$E \subseteq V \times \Gamma \times V$	 the set of edges, and
${\labl} : V \rightarrow \Sigma$   the node labelling.
The set of all graphs over $\Sigma$ and $\Gamma$ is denoted by
$\GR\Sigma\Gamma$.
We allow graphs that have both labelled and unlabelled
nodes and edges by introducing a designated  symbol $*$ to represent
an `unlabel' in our specifications, but we  omit
this symbol from our drawings.
We  write $\GR*\Gamma$ and $\GR\Sigma*$ to distinguish the
cases when all nodes are unlabelled, and all edges are unlabelled,
respectively.

\paragraph{Logic for graphs.}
For alphabets $\Sigma$ and $\Gamma$,
the monadic second-order logic $\MSO\Sigma\Gamma$
expresses properties of graphs over $\Sigma$ and $\Gamma$.
The logical language uses both node variables
$x, y, \ldots$  and node-set variables
$X, Y, \ldots $.

There are four types of atomic formulas:
$\lab_\sigma(x)$,
meaning node $x$ has label $\sigma$ (with $\sigma\in\Sigma$);
$\edge_\gamma(x,y)$,
meaning there is an edge from $x$ to $y$ with label $\gamma$
(with $\gamma \in\Gamma$);
$x=y$, meaning nodes $x$ and $y$ are equal; and
$x\in X$, meaning $x$ is an element of $X$.

As usual,
formulas are built from atomic formulas with the propositional connectives
$\neg, \wedge, \vee, \rightarrow$,
using the quantifiers $\forall$ and $\exists$
both for node variables and node-set variables.

A useful example \cite{thwr} of such a formula is
the binary predicate $\pat$ claiming the existence of a (directed)
path from $x$ to $y$:
\[ x\pat y \;=\; (\forall X) [ ( x\in X \wedge \closed(X) )
		   \rightarrow y\in X ]
\]
where
$ \closed(X) =
		   (\forall z_1)(\forall z_2)
		       (z_1\in X \wedge \edge(z_1,z_2) \rightarrow z_2\in X)
$,
and
$ \edge(z_1,z_2) = \bigvee_{\gamma\in\Gamma}\edge_\gamma(z_1,z_2) $.
We  also use $x\patuneq y$,
where one additionally requires that $x\neq y$;
for acyclic graphs this expresses the existence of a nonempty path
from $x$ to $y$.
\medskip

Let $\varphi$ be a formula of $\MSO\Sigma\Gamma$ with set $\Xi$ of
free variables (of either type),
and let $g=(V,E,\labl)$ be a graph in $\GR\Sigma\Gamma$.
Let $\nu$ be a valuation of $\varphi$, i.e.,
a mapping that assigns
to each node variable $x\in\Xi$ an element $\nu(x)$ of $V$, and
to each set variable $X\in\Xi$ a subset
$\nu(X)$ of $V$.
We write $g,\nu \models \varphi$ if $\varphi$ is satisfied in the graph $g$,
where the free variables of $\varphi$ are valuated according to $\nu$.

Let
$\varphi(x_1,\dots,x_m,X_1,\dots,X_n)$ be an $\MSO\Sigma\Gamma$  formula
with free node variables $x_i$
and free node set variables $X_j$,
and let $u_1,\dots,u_m$ be nodes of graph $g$,
and $U_1,\dots,U_n$ sets of nodes of $g$.
We write
$g\models\varphi(u_1,\dots,u_m,U_1,\dots,$ $U_n)$ whenever
$g,\nu\models \varphi(x_1,\dots,x_m,X_1,\dots,X_n)$,
where $\nu$ is the valuation with $\nu(x_i)=u_i$,
$\nu(X_j)=U_j$.

Let $\Xi$ be a finite set of variables.
The set $\pow{\Xi}$
of $0,1$-assignments to elements of $\Xi$ is finite, and
may be considered as  an alphabet.
A $\Xi$-{\em valuated} graph over $\Sigma$ and $\Gamma$
is a graph in $\GR{\Sigma\times\pow{\Xi}}\Gamma$,
such that for every node variable $x$ in $\Xi$ there is a unique node
of the graph of which the label $(\sigma,f)\in\Sigma\times\pow{\Xi}$
satisfies $f(x)=1$.

Clearly, such a $\Xi$-valuated graph  $g$
determines a graph $g\restrict\Sigma$ in
$\GR\Sigma\Gamma$,
by dropping the $\pow{\Xi}$ component of its node labels,
as well as a valuation $\nu_g$ of the variables in $\Xi$,
by taking

\begin{itemize}
\item[--]
for a node variable $x\in \Xi$,
$\nu_g(x) = u$, where $u$ is the unique node having a label $(\sigma,f)$
with $f(x)=1$,
\item[--]
for a node-set variable $X\in\Xi$,
$\nu_g(X) = U$, where $U$ consists of all nodes $v$ having a label $(\sigma,f)$
with $f(X)=1$.
\end{itemize}

For a formula $\varphi$ of $\MSO\Sigma\Gamma$ with free variables in $\Xi$,
and a $\Xi$-valuated graph $g$
we write $g\models\varphi$ if
$\varphi$ is true for the underlying graph under the implicitly defined
valuation, i.e., if
$g\restrict\Sigma, \nu_g \models \varphi$;
$\varphi$ defines the graph language $GL(\varphi)=
\{ g \in \GR{\Sigma\times\pow{\Xi}}\Gamma \mid g\models\varphi \}$.
A graph language is {\em mso definable} if there exists a closed mso formula
that defines the language.

\paragraph{String representation.}
A  string $w\in\Sigma^*$ of length $k$ can
be represented	by the graph $\ngr(w)$ in $\GR\Sigma*$,
consisting of $k$ nodes labelled by the consecutive symbols of $w$, with
$k-1$ (unlabelled) edges representing the successor relation for the
positions of the string.
In the figure below, we show $\ngr(ababb)$.
Note that for the empty string $\empt$, $\ngr(\empt)$ is the empty graph.
With this representation,
a formula $\varphi$ of $\MSO\Sigma*$ defines the string language
$L(\varphi) = \{ w\in(\Sigma\times\pow\Xi)^* \mid \ngr(w) \models \varphi \}$,
where $\Xi$ is the set of free variables of $\varphi$;
note that $\ngr(w)$ is a $\Xi$-valuated graph over $\Sigma$ and $*$.

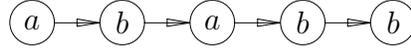
\centerline{\unitlength 0.60mm
\begin{picture}(140,20)
\input 2wgsmvb.gtp
\end{picture}
}

Given the close connection between the positions and their
successor relation in a string $w$ on the one hand,
and the nodes and their connecting  edges in $\ngr(w)$ on the other,
we  say that a string $w$ satisfies a formula $\varphi$
if $\ngr(w) \models \varphi$.

String languages definable by monadic second-order formulas are exactly the
regular languages, as shown by B\"uchi and Elgot.

\begin{prop}[\cite{buchi,elgot}]\label{buchi}
\leavevmode

\begin{enumerate}
\item
   $L(\varphi)$ is a regular string language
   for every formula $\varphi$ of $\MSO\Sigma*$.
\item
   A string language $K\subseteq\Sigma^*$ is regular iff there is a
  closed formula $\varphi$ of $\MSO\Sigma*$ such that $K=L(\varphi)$.
\end{enumerate}
\end{prop}

We will also refer to Proposition~\ref{buchi} as `B\"uchi's result',
with due apologies to Elgot.

   Observe that the set of all strings over a fixed alphabet
$\Sigma$ forms an mso definable {\em graph} language via the above
representation.
The defining formula
for the set $\{\ngr(w) \mid w\in\Sigma^* \}$
over $\MSO\Sigma*$
expresses the existence of an initial and a final node
(provided the graph is nonempty)
and demands that every node has at most one direct successor
(i.e., the edge relation is functional);
`guards'  $(\exists x)\true \to$ are added
in order to make the empty string $\empt$ satisfy the formula.
\begin{eqnarray*}
& &
(\exists x)\true \to
(\exists x)(\forall y)(x\pat y \land \neg(y\patuneq x )) \\
& \land
&
(\exists x)\true \to
(\exists x)(\forall y)(y\pat x \land \neg(x\patuneq y )) \\
& \land
& (\forall x)(\forall y_1)(\forall y_2)
((\edge(x,y_1) \land \edge(x,y_2)) \to y_1=y_2)
\end{eqnarray*}

As a consequence, the set of graphs representing a 
string language $K$,
$\{\ngr(w) \mid w\in K \}$ is an mso definable graph language 
for every regular language $K$.

\section{Two-Way Machines}\label{sect:2way}

We present our (slightly nonstandard) model of
two-way generalized sequential machines (2gsm),
or two-way finite state transducers.
In order to facilitate the proof of the equivalence of two-way
finite state transductions and logically definable transductions
we extend the basic model to a machine model that has its input tests
as well as moves specified by mso formulas.
We prove the equivalence of this extended model to the basic model.
An important tool in this proof is the observation that a two-way
automaton is able to keep track of the state of another (one-way)
finite state automaton
(proved in Lemma~3 of \cite{houl67}, see also p.~212 of \cite{ahu69}).
We formalize this fact by extending the 2gsm with the feature of
`regular look-around'.
The equivalence of this model with the basic model is then proved using
the related result of
\cite{chytil} stating that deterministic two-way
finite state transductions are closed under composition.
The equivalence of the regular look-around model with the
mso formula model is proved using B\"uchi's result
(Proposition~\ref{buchi}).

\medskip
Since we need several types of two-way machines,
we first introduce a  generic model,
and then instantiate it in several ways.
\medskip

A {\em two-way	machine} (2m)
is a finite state device
equipped with a two-way input tape (read only), and a one-way output tape.
In each step of a computation the machine
  reads an input symbol,
  changes its internal state,
  outputs a string,
and
  moves its input head,
all depending on the symbol read and the original internal state.


We specify a 2m as a construct
  $\cM = (Q,\Sigma_1,\Sigma_2,\delta,q_\ini,q_f)$,
where
  $Q$ is the finite set of states,
  $\Sigma_1$ and $\Sigma_2$ are the input alphabet and output alphabet,
  $q_\ini$ and $q_f$ are the initial and the final state,
and
  $\delta$ is a finite set of {\em instructions}.
Each instruction is of the form
  $(p,t, \; q_1,\alpha_1,\mu_1, \; q_0,\alpha_0,\mu_0)$,
where $p\in Q-\{q_f\}$ is 
the present state of the machine,
  $t$ is a test to be performed on the input,
and the triples
  $(q_i,\alpha_i,\mu_i)$, $i=1,0$,
fix the action of the machine depending on the outcome of the test $t$:
  $q_i\in Q$ is the new state,
  $\alpha_i\in \Sigma_2^*$ is the string written on the output tape, and
  $\mu_i$ describes the (deterministic) move
of the reading head on the input tape.
The precise form of these instructions varies from one model to another,
in particular the form of the test $t$, and the moves $\mu_i$.

The above instruction can be expressed as the following informal code.

\begin{minipage}{\textwidth}
\begin{tt}\small
\begin{tabbing}
label $p$: \= if $t$ \= then write $\alpha_1$ ; move $\mu_1$ ; goto $q_1$  \\
	   \>	     \> else write $\alpha_0$ ; move $\mu_0$ ; goto $q_0$  \\
	   \> fi
\end{tabbing}
\end{tt}
\end{minipage}

\smallskip
The string on the input tape is marked by two special symbols,
$\ltape$ and $\rtape$, indicating the boundaries of the tape.
So, when processing the string $\sigma_1\cdots \sigma_n$,
$\sigma_i\in \Sigma_1$,
the tape has $n+2$ reachable positions $0,1,\dots,n,n+1$,
containing the string	  $\tape{\sigma_1\cdots \sigma_n}$.
The reading head is on one of these positions.

The 2m $\cM$ realizes the transduction
$m\subseteq \Sigma_1^*\times\Sigma_2^*$,
such that $(w,z)\in m$ whenever there exists a computation with $\tape{w}$
on the input tape, starting in initial state $q_\ini$ with the input
head on position $0$ (where the symbol $\ltape$ is stored), and
ending in the accepting state $q_f$,
while $z$ has been written on the output tape.
\medskip

A 2m is {\em deterministic} if for each state $p$
there is at most one instruction
   $(p,t,\; q_1,\alpha_1,\mu_1,\; q_0,\alpha_0,\mu_0)$
that starts in $p$.
Note that the transduction $m$ realized by a deterministic 2m $\cM$
is a partial function $m: \Sigma_1^* \to \Sigma_2^*$
because the $\mu_i$ in the instructions describe deterministic moves of the
reading head.

\medskip
We  consider the usual two-way {\em generalized sequential} machine
(2gsm), introduced in \cite{ahul70},
 and two new instantiations of the generic 2m model,
the 2gsm with {\em regular look-around}, and
the 2gsm with {\em mso-instructions}.


\paragraph{2gsm.}
For the basic 2gsm model each instruction
  $(p,t, \; q_1,\alpha_1,\mu_1, \; q_0,\alpha_0,\mu_0)$ in $\delta$
satisfies
  $t\in \Sigma_1\cup\{\ltape,\rtape\}$,
and
  $\mu_i\in \{-1,0,+1\}$, $i=1,0$.

Executing an instruction
   $(p,\sigma,\; q_1,\alpha_1,\epsilon_1,\; q_0,\alpha_0,\epsilon_0)\in\delta$
the 2gsm,
assuming it is in internal state $p$,
when reading $\sigma$ on its input tape,
changes its state to $q_1$,
writes $\alpha_1$ to its output tape,
and moves its head from the present position $i$ to
the position $i+\epsilon_1$ (provided $0\leq i+\epsilon_1\leq n+1$);
if $\sigma$ is not read on the input tape it acts similarly according
to the triple
   $(q_0,\alpha_0,\epsilon_0)$.
Recall that there are no instructions starting in the final state.
\smallskip

It is more customary to formalize the instructions of a 2gsm as
5-tuples
   $(p,\sigma,q,\alpha,\epsilon)$,
not having the `else-part' of our instructions.
These two approaches are easily seen to be equivalent.
Obviously, the 5-tuple can be extended to an 8-tuple
by adding a dummy `else-part', as in
   $(p,\sigma,\;	q,\alpha,\epsilon,\;  p,\empt,0)$.
Conversely,
one of our instructions
   $(p,\sigma,\; q_1,\alpha_1,\epsilon_1,\; q_0,\alpha_0,\epsilon_0)$
can be replaced by the `if-part'
   $(p,\sigma, q_1,\alpha_1,\epsilon_1)$
and all alternatives
   $(p,\sigma', q_0,\alpha_0,\epsilon_0)$,  $\sigma'\neq\sigma$.

For determinism we require each state to have at most one instruction,
whereas the customary notion considers both state and input symbol.
This, somewhat unusual, formulation allows us to have 
the above common definition
of determinism for all necessary instantiations of our generic model,
without having to worry about the mutual exclusiveness of the tests $t$.
This is the 
reason for choosing our 8-tuple formalism.

The first of the two translations (from 5-tuple model to our 8-tuple model)
does not respect determinism.
We can solve this by checking all alternatives in a given state consecutively,
as follows.
Let
   $(p,\sigma_i,q_i,\alpha_i,\epsilon_i)$,   $i=1,\dots,k$
be all the instructions for state $p$ in a deterministic (5-tuple) 2gsm,
which means that the $\sigma_i$ are different.
Introduce $k+1$ copies $p=p^{(1)}, p^{(2)}, \dots,p^{(k)}, p^{(k+1)}$ of $p$.
Then, the instructions
   $(p^{(i)},\sigma_i,\; q_i ,\alpha_i,\epsilon_i,\;  p^{(i+1)},\empt,0)$,
   $i=1,\dots,k$,
offer the same alternatives, but sequentially rather than in parallel.

\begin{exple}\label{ex1}
Consider the string transduction

\[\tradu.\]

An obvious deterministic
2gsm reads each segment of $a$'s from left to right while
copying it to the output.
When encountering
a $b$ it rereads the segment from right to left. 
This second pass it writes $b$'s to the output tape.

This machine can be implemented by taking
$\Sigma_1=\Sigma_2=\{a,b\}$,
$Q=\{ 0,1,2,3,4,5 \}$,
$q_\ini=0$, $q_f=5$,
and
$\delta$ consisting of the instructions

\begin{Itemize}
\item[]
$( 0, \ltape,  \; 1, \empt, +1, \;  0, \empt, 0  )$
\item[]
$( 1, a,       \; 1, a, +1,	\; 2, \empt, 0	)$
\item[]
$( 2, b, \;  3, \empt, -1, \;  5, \empt, 0 )$
\item[]
$(3, a,   \; 3, b, -1,	\;   4, \empt, +1 )$
\item[]
$( 4, a, \;  4, \empt, +1,  \;	1, \empt, +1)$
\end{Itemize}

Note that the last three elements of the first instruction
are irrelevant.

The computation of the 2gsm on input $aaabbaba$
can be visualized as in
Figure~\ref{fig1}, where we have labelled the edges of the computation
by the strings that are written to the output
(with $\empt$ omitted, for convenience).
\end{exple}

\begin{figure}[tbh]
\centerline{\unitlength 0.60mm
\begin{picture}(20,100)
\input 2n-cmpl.gtp
\end{picture}%
\begin{picture}(20,100)
\input 2n-cmpaa.gtp
\end{picture}%
\begin{picture}(20,100)
\input 2n-cmpaa.gtp
\end{picture}%
\begin{picture}(20,100)
\input 2n-cmpaa.gtp
\end{picture}%
\begin{picture}(20,100)
\input 2n-cmpab.gtp
\end{picture}%
\begin{picture}(20,100)
\input 2n-cmpbb.gtp
\end{picture}%
\begin{picture}(20,100)
\input 2n-cmpba.gtp
\end{picture}%
\begin{picture}(20,100)
\input 2n-cmp0b.gtp
\end{picture}%
\begin{picture}(20,100)
\input 2n-cmp0a.gtp
\end{picture}%
\begin{picture}(20,100)
\input 2n-cmpr.gtp
\end{picture}%
}
\caption{Computation for $(a^3b^2aba, a^3b^3aba)$
	 of 2gsm from Example~\ref{ex1}}\label{fig1}
\end{figure}


\paragraph{Look-around.}
A {\em 2gsm with regular look-around} (2gsm-rla) extends the basic
2gsm model, by allowing more complicated tests.
In an instruction
   $(p,t, \; q_1,\alpha_1,\epsilon_1, \; q_0,\alpha_0,\epsilon_0)\in\delta$
all components are as before for the 2gsm,
except the test $t$, which does not consist of a single letter $\sigma$,
but of a triple $t=(R_\ell,\sigma,R_r)$,
where $\sigma\in (\Sigma_1\cup\{\ltape,\rtape\})$,
and $R_\ell,R_r$ are regular languages such that
$R_\ell, R_r \subseteq (\Sigma_1\cup \{\ltape, \rtape\})^*$.
This test $t$ is satisfied if $\sigma$ is the symbol under the reading
head, and the strings to the left and the right of the head belong to
$R_\ell$ and $R_r$ respectively.

Obviously, it suffices to have tests $(R_\ell,\sigma,R_r)$
such that
 $R_\ell \cdot \sigma \cdot R_r \subseteq \ltape \Sigma_1^*\rtape $.
For a given 2gsm-rla, an equivalent 2gsm-rla with that property is obtained
by changing each test $(R_\ell,\sigma,R_r)$ into
$(R'_\ell,\sigma,R'_r)$ where
$R'_\ell = R_\ell \cap \ltape\Sigma_1^*$
(with the exception that $R'_\ell = \{\empt\}$ when $\sigma=\ltape$),
and similarly for $R'_r$.
We observe here that this notion of `regular look-around'
generalizes the well-known notion of regular look-ahead for one-way
automata (see, e.g., \cite{nijholt,eng77}).


\paragraph{Mso instructions.}
For a {\em 2gsm with mso-instructions} (2gsm-mso) the test and the moves
of each instruction are given by mso formulas.
To be precise, for
$(p,t,\; q_1,\alpha_1,\mu_1, \; q_0,\alpha_0,\mu_0)\in\delta$,
$t$ is given as a formula $\varphi(x)$
in  $\MSO{\Sigma_1\cup\{\ltape,\rtape\}}*$
with one free node variable $x$, and the moves $\mu_i$ are given by
functional formulas $\varphi_i(x,y)$
in  $\MSO{\Sigma_1\cup\{\ltape,\rtape\}}*$
with two free node variables $x$ and $y$
(see below for the meaning of `functional').

A test $t=\varphi(x)$ is evaluated for the string
on the input tape
with $x$ valuated as the position taken by the reading head;
more precisely,
as our logic is defined for graphs, $t$ is true whenever
$\ngr(\tape{w}) \models \varphi(u)$,
where $w$ is the input string, and $u$ is the node corresponding to
the position of the reading head.

The 2gsm-mso does not move step-wise on the input tape,
but it `jumps' as specified by the formulas
$\varphi_i(x,y)$, as follows.
Assuming the machine is in position $u$, it moves to a position $v$
for which
$\ngr(\tape{w}) \models \varphi_i(u,v)$,
where we have identified positions on the input tape with
their corresponding nodes of the graph
$\ngr(\tape{w})$.

To guarantee that the $\varphi_i(x,y)$ describe deterministic moves
of the reading head,
we require that the relations specified by
$\varphi_i(x,y)$ are functional,
for each input string $w$,
i.e., for every position $u$ there is at most one position $v$
such that $\ngr(\tape{w}) \models \varphi_i(u,v)$.
Note that
functionality is expressible in the logic:
$(\forall x)(\forall y_1)(\forall y_2)
       [\; \varphi_i(x,y_1) \land \varphi_i(x,y_2) \to y_1=y_2 \;]$.
Consequently, it is decidable;
we may use B\"uchi's result
(Proposition~\ref{buchi}, which is effective)
to verify that it is satisfied by every string in $\tape{\Sigma_1^*}$.

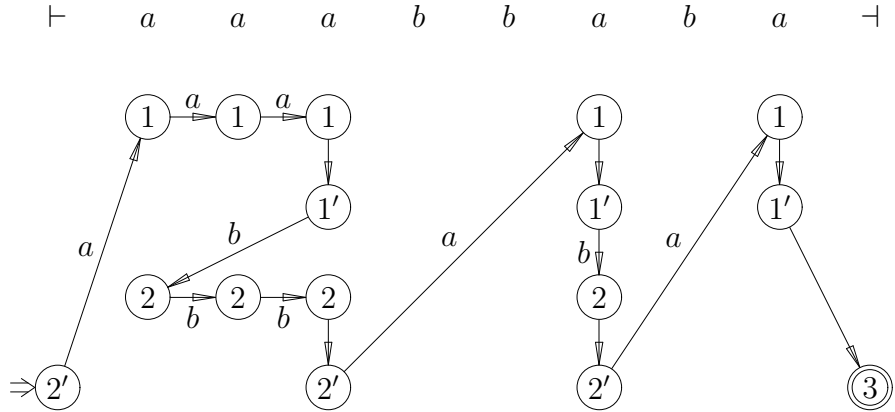
\begin{figure}[tb]
\centerline{\unitlength 0.60mm
\begin{picture}(180,100)(0,20)
\input 2n-2mso2.gtp
\end{picture}%
}
\caption{Computation for $(a^3b^2aba, a^3b^3aba)$
	 of 2gsm-mso from Example~\ref{ex1a}}\label{fig:ex1a}
\end{figure}

\begin{exple}\label{ex1a}
Consider again the string transduction
$m=$

\[\tradu.\]

We use the predicate $\mbox{next}_a(x,y)$
to specify the first position $y$ following $x$ that is labelled by $a$:
\[ x\patuneq  y \land  \lab_a(y) \land
   (\forall z)\left[\;
	      ( x\patuneq z \land  z \patuneq y ) \to \neg
       \lab_a(z) \; \right]
\]

Similarly we construct an expression
$\form{fis}_a(x,y)$
denoting the first $a$ in the present segment of $a$'s,
\[
y \pat x \land
	     (\forall z)( y \pat z \land z \pat x \to \lab_a(z) )
	  \land
	     \neg(\exists z)( \edge_*(z,y)  \land     \lab_a(z) )
\]

Using these predicates we build a deterministic 2gsm-mso that realizes $m$.
In state $1$ it walks along a segment of $a$'s, copying it to the
output tape. Then, when the segment is followed by a $b$, it jumps back
to the first $a$ of the segment for a second pass, in state $2$.
When the end of the segment is reached for the second time,
the machine jumps to the next segment, returning to state $1$.
At the last $a$ of the input the machine jumps to the right end marker,
and halts in the final state $3$.

Let
$\Sigma_1=\Sigma_2=\{a,b\}$,
$Q=\{ 1,1',2,2',3 \}$,
$q_\ini=2'$, $q_f=3$,
and
$\delta$ consisting of the transitions

\begin{Itemize}
\item[]
$( 1, (\exists y)( \edge_*(x,y) \land \lab_a(y)  ), \;
   1, a, \edge_*(x,y), \;
   1', \empt, x=y )
$
\item[]
$( 1', (\exists y)( \edge_*(x,y) \land \lab_b(y)  ), \;
   2, b, \form{fis}_a(x,y), \;
   3, \empt, \lab_\rtape(y)  )
$
\item[]
$( 2, (\exists y)( \edge_*(x,y) \land \lab_a(y)  ), \;
   2, b, \edge_*(x,y), \;
   2', \empt, x=y )
$
\item[]
$( 2', (\exists y)( x \patuneq y  \land \lab_a(y)  ), \;
   1, a, \mbox{next}_a(x,y)  , \;
   3, \empt, \lab_\rtape(y)  )
$
\end{Itemize}

The computation of the machine on  input $a^3b^2aba$ can be visualized as in
Figure~\ref{fig:ex1a}
(where, again, $\empt$ is omitted from the edges of the computation).
\end{exple}

Without loss of generality we assume that the 2m's we consider
never write  more than one symbol at a time,
i.e., for each instruction
   $(p,\sigma,\; q_1,\alpha_1,\mu_1,\;$ $q_0,\alpha_0,\mu_0)$
we have
   $|\alpha_i| \le 1$  (for $i=1,0$).

\medskip
We abbreviate deterministic 2m's by adding a `d' to the usual
abbreviation, hence we speak of 2dgsm, 2dgsm-rla, and 2dgsm-mso.
The families of string transductions realized by these three types of
deterministic sequential machines are denoted by
$\DGSM$, $\DSMrla$, and $\DSMmso$,
respectively.

Unlike their nondeterministic counterparts
(\cite{kiel}, see also Lemma~\ref{lem:not-gsm}
and the remark following it),
deterministic 2gsm's are closed under composition, as was
demonstrated by Chytil and J\'akl.
As an essential part of the proof the fact is used
(proved in \cite{houl67})
that a 2dgsm can keep track of the state of another
(deterministic) one-way finite state automaton working on the same
tape (from left to right or from right to left).
For the left-to-right case,
it is clear how to do this as long as the reading head moves to the
right.
Backtracking (`undoing' a move) on the occasion of a step to the left,
needs a rather ingenious back and forth simulation of the automaton.

\begin{prop}[\cite{chytil}]\label{prop:2dgsm-composition}
   $\DGSM$ is closed under composition.
\end{prop}

In the remainder of this section we show that the three types of deterministic
machines defined above are all equivalent,
i.e., that
$\DGSM	= \DSMrla = \DSMmso $.
\smallskip

Every 2gsm is of course a simple 2gsm-rla,
using trivial look-around tests,
i.e., tests of the form
$(R_\ell,\sigma,R_r)$,	with
$R_\ell = \ltape \Sigma_1^*$, and
$R_r = \Sigma_1^* \rtape$
(with the exceptions $R_\ell = \{\empt\}$ when $\sigma=\ltape$,
 and $R_r = \{\empt\}$ when $\sigma=\rtape$).

\smallskip
It follows from B\"uchi's result, Proposition~\ref{buchi},
that any 2gsm-rla can be reinterpreted as a 2gsm-mso
by changing the specification of the tests and moves into formulas,
as follows.

First, consider a look-around test
$t=(R_\ell,\sigma,R_r)$.
Let $\psi_\ell(x)$ be a formula expressing that the string to the left of
position $x$ belongs to the regular language $R_\ell$.
It can be obtained from a closed formula $\psi$ defining $R_\ell$
by restricting
quantification to the positions to the left of $x$,
i.e., by replacing subformulas
$(\exists y)\xi(y)$  by  $(\exists y)(y\patuneq x \land \xi(y))$
and
$(\exists Y)\xi(Y)$  by
$(\exists Y)( (\forall y)(y\in Y \to y\patuneq x) \land \xi(Y))$.

Similarly, we obtain a formula $\psi_r(x)$ expressing that
the string to the right of
position $x$ belongs to the regular language $R_r$.
Clearly, the test $t$ is equivalent to the formula
$\varphi_t(x)= \psi_\ell(x)\land \lab_\sigma(x)\land \psi_r(x)$.

Finally,
one-step moves are easily translated into formulas.
A move $\epsilon=+1$ is equivalent to stating that the new position
is next to the original:
$\edge_*(x,y)$.
Of course, $\epsilon =-1$ is symmetric,
whereas $\epsilon =0$ is expressed by	$x=y$.
Note that these formulas are functional.
\medskip

These observations prove the first relations between
the families of transductions.

\begin{lem}
$\DGSM	\subseteq \DSMrla \subseteq \DSMmso $.
\end{lem}

The feature of 2dgsm's that they can keep track of the state of a
one-way finite state automaton
(cf. the remark before Proposition~\ref{prop:2dgsm-composition}),
is modelled by us as regular look-around.
Thus, for readers familiar with this feature it should be quite
obvious that $\DSMrla \subseteq \DGSM$.
Here we prove it using Proposition~\ref{prop:2dgsm-composition}.

\begin{lem}\label{rla-dgsm}
$\DSMrla \subseteq \DGSM$.
\end{lem}

\begin{proof}
By Proposition~\ref{prop:2dgsm-composition},
$\DGSM$ is closed under composition. We  prove the lemma
by decomposing a given 2dgsm-rla $\cM$ into a series of 2dgsm's,
together realizing  the transduction of $\cM$.

The final 2dgsm  performs the required transduction,
whereas all the other transductions  `preprocess the tape',
by adding to the original input
the outcome of the various tests of $\cM$.
As we also need this information for the positions containing
the end-of-tape markers \ltape\ and \rtape,
we start by a transduction that maps input $w$ to the string
$\ininode w \finnode$,
where \ininode\ and \finnode\ are new symbols.
Information concerning the end-of-tape positions is added to these
new symbols.
The other machines may ignore \ltape\ and \rtape,
and treat \ininode\ and \finnode\ as if they where these end-of-tape markers.

\smallskip
For each look-around test $t=(R_\ell,\sigma,R_r)$ of $\cM$
we introduce a 2dgsm $\cM_t$ that copies the input, while adding
to each position the outcome of the test $t$ for that position
in the original string (ignoring any other additional information
a previous transduction added to the string).
The machine
$\cM_t$ itself can be seen as the work of three consecutive 2dgsm's.
The first one, simulating a finite state automaton recognizing $R_\ell$,
checks on each position whether the prefix read belongs to $R_\ell$.
It adds this information to the symbol at that position.
The second transducer, processing the input from right to left,
simulating a finite state automaton for the mirror image of $R_r$,
adds information concerning the suffix.
Note that the input has been reversed in the process.
This can be undone by another reversal performed by a third 2dgsm.

\smallskip
Once the value of each look-around test of $\cM$ is added to the original
input string,
obviously the transduction of $\cM$ can be simulated by an ordinary 2dgsm.
\end{proof}

B\"uchi's result (Proposition~\ref{buchi})
allows us to show that the 2gsm-mso can be simulated by the 2gsm-rla.
Additionally we need the following (folklore) result on the structure of
certain regular languages
(cf.  \cite[Lemma~8.1]{pixton}).

\begin{lem}\label{cdot}
Let $\Delta\subseteq\Sigma$ be alphabets, and let
$R\subseteq\Sigma^*$ be a regular language such that each string of $R$
contains exactly one occurrence of a symbol from $\Delta$.
Then we may write $R$ as a finite union of disjoint languages
$R_\ell\cdot a \cdot R_r$,
where $a\in\Delta$, and $R_\ell, R_r \subseteq (\Sigma-\Delta)^*$
are regular languages.
\end{lem}

\begin{proof}
Let $\cA$ be a deterministic finite automaton accepting $R$.
Every path (in the state transition diagram of $\cA$)
from the initial state to a final state passes exactly one transition
labelled by a symbol from $\Delta$.
For any such transition $(p,a,q)$ of $\cA$ let
$R_\ell$ consist of all strings that label a path starting in the initial
state of $\cA$ and ending in $p$, and symmetrically, let
$R_r$ consist of all strings that label a path from $q$
to one of the final states of $\cA$.
Obviously, $R_\ell$ and $R_r$ are regular, and $R$
is the union of the languages
$R_\ell\cdot a\cdot R_r$ taken over all such transitions.
Since $\cA$ is deterministic, these languages are easily seen to
be disjoint.
\end{proof}

\begin{lem}
$\DSMmso \subseteq \DSMrla$.
\end{lem}

\begin{proof}
We show how to simulate the instructions of a 2gsm-mso
by a 2gsm-rla.
Recall that such an instruction is specified as
   $(p,t, \; q_1,\alpha_1,\mu_1, \; q_0,\alpha_0,\mu_0)$,
where
 $t$ is a formula $\varphi(x)$ with one free node variable,
and the
 moves $\mu_i$ are  (functional)
formulas $\varphi_i(x,y)$ with two free node variables.

\medskip\noindent{\it Tests: unary node predicates.}
Consider a test $\varphi(x)$ in $\MSO{\Sigma_1\cup\{\ltape,\rtape\}}*$.
It can easily be simulated by regular look-around tests.
Identifying  $(\Sigma_1\cup\{\ltape,\rtape\}) \times\pow{\{x\}}$ with
$(\Sigma_1\cup\{\ltape,\rtape\}) \times\pow{}$,
consider the  language $L(\varphi)$,
which is regular by Proposition~\ref{buchi}.
As each string of this language contains exactly one symbol with $1$ as its
second component,
it can be written as a finite union of languages
$R_\ell \cdot (\sigma,1) \cdot R_r$,
with regular languages
$R_\ell, R_r \subseteq ((\Sigma_1\cup\{\ltape,\rtape\}) \times\{0\})^*$,
and
$\sigma\in\Sigma_1\cup\{\ltape,\rtape\}$,
see Lemma~\ref{cdot}.
This implies that the test $\varphi(x)$ can be simulated by a finite disjunction
of the look-around tests  $(R'_\ell, \sigma, R'_r)$,
where each $R'_\ell, R'_r$ is obtained from the corresponding
$R_\ell, R_r$ by dropping the second component (the 0-part) of the symbols.
Of course, this disjunction is computed by testing each of its alternatives
consecutively.

\medskip\noindent{\it Moves: binary node predicates.}
Once the test of an instruction is evaluated, one of its moves is executed,
and the output is written.
This move is given as a  formula $\varphi(x,y)$,
specifying a functional relation
between the present position $x$ and the next position $y$ on the input.
Where the 2dgsm-mso may `jump' to its next position,
independent of the relative positions of $x$ and $y$,
a 2dgsm-rla can only step to one of the neighbouring positions of the tape,
and has to `walk' to the next position when simulating this jump.

Before starting the excursion from $x$ to $y$ the 2dgsm-rla determines the
direction (left, right, or stay) by evaluating the tests
$(\exists y)(y\prec x \land \varphi(x,y))$,
$(\exists y)(x\prec y \land \varphi(x,y))$, and
$(\exists y)(x=y \land \varphi(x,y))$
using the method that we have explained above.
Since $\varphi(x,y)$ is functional, at most one of these tests is true.

In the sequel we assume that our target position $y$ lies to the left
of the present position $x$, i.e.,
test
$(\exists y)(y\prec x \land \varphi(x,y))$ is true.
The right-case can be treated in an analogous way;
the stay-case is trivial.

Similarly to the case of tests,
identify  $(\Sigma_1\cup\{\ltape,\rtape\}) \times\pow{\{x,y\}}$ with
$(\Sigma_1\cup\{\ltape,\rtape\}) \times\pow{2}$, and
consider $L(y\prec x \land\varphi(x,y))$.
Each string of this language contains exactly one symbol with $(0,1)$ as its
second component, the position of $y$, and it precedes a unique symbol
with $(1,0)$ as its second component, the position of $x$;
all other symbols carry $(0,0)$.
It can be written as a finite disjoint
union of languages
$R_\ell \cdot (\sigma,0,1) \cdot R_m \cdot (\tau,1,0) \cdot R_r$,
with regular languages
$R_\ell, R_m, R_r \subseteq
  ((\Sigma_1\cup\{\ltape,\rtape\}) \times\{(0,0)\})^*$
  and $\sigma, \tau \in \Sigma_1\cup \{\ltape, \rtape\}$,
by applying Lemma~\ref{cdot} twice.

Our moves are functional,
meaning that there is a unique position
$y$ that satisfies the predicate $\varphi(x,y)$ with $x$ the present position.
Still before starting the excursion from $x$ to 
the new position $y$, the 2dgsm-rla
 determines which  language in the union above
describes this position
by performing the regular look-around tests
$(R'_\ell\cdot \sigma \cdot R'_m,  \tau, R'_r)$,
where each $R'_\ell, R'_m, R'_r$ is obtained from the corresponding
$R_\ell, R_m, R_r$ by deleting the second component (the (0,0)-part)
of the symbols.

The 2dgsm-rla now moves to the left.
In each step it checks whether the segment
of the input string between the present position (candidate $y$)
and the starting position (corresponding to $x$)
belongs to the regular language $R'_m$.
This can be done by simulating a finite automaton for (the mirror image of)
$R'_m$ in the finite state control.

Each time this segment belongs to $R'_m$,
it performs the rla-test
$(R'_\ell, \sigma, \Sigma_1^*\rtape)$,
to verify the requirement on the initial segment of the input.
Once this last test is satisfied, it has found the position $y$
 and writes the output string.
\end{proof}

We summarize.

\begin{thm}\label{dgsm=dgsm-mso}
$\DGSM=\DSMrla=\DSMmso$.
\end{thm}

A similar result can be obtained for nondeterministic gsm's
by the same line of reasoning. 
However, in Lemma~\ref{rla-dgsm}
we need the inclusion $\DGSM\cmp\NGSM \subseteq \NGSM$
rather than $\DGSM\cmp\DGSM \subseteq \DGSM$
(Proposition~\ref{prop:2dgsm-composition}).
This new inclusion can be proved like the latter one \cite{chytil}.


\section{MSO Definable String Transductions}\label{sect:mso}

As explained in the Preliminaries,
we consider mso logic on graphs as a means of specifying
string transductions,
rather than dealing directly with strings.

Although     we are mainly interested in graph transductions that
have string-like graphs as their domain and range,
occasionally we find it useful to allow more general
graphs as intermediate products of our constructions.

In this section we recall the definition of mso graph transductions,
and from it we derive two families of mso definable string transductions,
which differ in the way strings are represented by graphs.
We present basic examples,
and characterize the relation between the two families we have defined.

\medskip
We start with the general definition.

\medskip
An {\em mso definable transduction}
\cite{msoV,msotrans,joost,engooslog,seese}
 is a (partial) function that constructs
for a given input graph a new output graph
as specified by a number of mso formulas.
Here we consider the deterministic (or, `parameterless') mso transductions
of \cite{msotrans}.
For a graph satisfying a given domain formula $\varphi_\dom$
we take copies of each of the nodes,
one for each element of a finite copy set $C$.
 The label of the $c$-copy of node
$x$ ($c\in C$) is determined by a set of formulas $\varphi_\sigma^c(x)$,
one for each symbol $\sigma$ in the output alphabet.
We keep only those copies of the nodes for which exactly one of the
label formulas is true.
Edges are defined according to formulas $\varphi_{\gamma}^{c_1,c_2}(x,y)$:
we construct an edge with label $\gamma$
in the output graph from the $c_1$-copy of $x$
to the $c_2$-copy of $y$
whenever such a formula holds.

\begin{defn}\label{def-mso}
An {\em mso definable (graph) transduction}
$\tau: \GR{\Sigma_1}{\Gamma_1} \rightarrow \GR{\Sigma_2}{\Gamma_2}$
is specified by

\begin{itemize}
\item[--]
       a closed {\em domain formula} $\varphi_{\dom}$,
\item[--]
       a finite {\em copy set} $C$,
\item[--]
      {\em node formulas} $\varphi_{\sigma}^{c}(x)$,
      with one free node variable $x$,
      for every $\sigma \in \Sigma_2$ and every $c \in C$, and
\item[--]
      {\em edge formulas} $\varphi_{\gamma}^{c_1,c_2}(x,y)$
      with two free node variables $x,y$,
 for every $\gamma \in \Gamma_2$ and all $c_1,c_2 \in C$,
\end{itemize}

where all formulas are in $\MSO{\Sigma_1}{\Gamma_1}$.

\medskip
\noindent For $g \in GL(\varphi_{\dom})$ with node set $V_g$, the image
$\tau(g)$ is the graph $(V, E, \labl)$,
defined as follows.
We will write $u^c$ rather than $(u,c)$ for elements of $V_g\times C$.

\begin{itemize}
\item[--]$\!\!
\begin{array}[t]{rl}
V =
	\{u^{c} \mid & u \in V_{g}, c\in C,\\ 
	& \mbox{there is exactly one }
   \sigma \in \Sigma_2 \mbox{ such that } g \models \varphi_\sigma^{c}(u)\},
\end{array}$
\item[--]
  $E = \{(u^{c_1},\gamma,v^{c_2}) \mid u^{c_1}, v^{c_2}\in V,
      \gamma\in \Gamma_2,
	g \models \varphi_{\gamma}^{c_1,c_2}(u,v) \}$, and
\item[--]
  $\labl(u^{c}) = \sigma$	if  $g\models \varphi^{c}_\sigma(u) $,
     for $u^{c} \in V$, $\sigma\in \Sigma_2$.
\end{itemize}
\end{defn}

\begin{exple}
Let $\Sigma=\{a,b\}$.
As a simple example we present an mso graph transduction
from  $\GR\Sigma*$ to $\GR*{\{a,b,*\}}$
that
transforms a linear graph representing a string
into a ladder, while moving the symbols from the nodes to the steps.

\begin{Itemize}
\item[]
Domain formula 
$\varphi_\dom$ expresses that the input graph is a string representation
(see the end of Section~\ref{preliminaries}).
\item[]
The copy set $C$ is $\{1,2\}$.
\item[]
Each node is copied twice:
$\varphi_*^1 = \varphi_*^2 = \true$.
\item[]
Unlabelled edges are copied twice,
one of these in reverse:	\\
$\varphi_*^{1,1} = \edge_*(x,y)$,
$\varphi_*^{2,2} = \edge_*(y,x)$,
$\varphi_*^{1,2} = \varphi_*^{2,1} = \false$.
\item[]
Labelled edges are introduced:\\
$\varphi_\sigma^{1,2} = (x=y) \land \lab_\sigma(x)$,
$\varphi_\sigma^{1,1} =
\varphi_\sigma^{2,1} = \varphi_\sigma^{2,2}  =\false$,
for $\sigma = a,b$.
\end{Itemize}

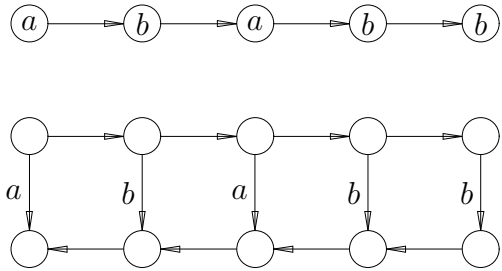
\centerline{\unitlength 0.50mm
\begin{picture}(120,85)
\input 2wgsmdub.gtp
\end{picture}
}
\end{exple}

The family of mso definable graph transductions is denoted by
\grMSO.
Its basic properties are summarized below,
see, e.g., \cite[Prop.~5.5.6]{cou97}.

\newpage
\begin{prop}\label{mso-closed}\leavevmode
\begin{enumerate}
\item
\grMSO\ is closed under composition.
\item
The mso definable graph languages are closed under inverse
mso definable graph transductions.
\end{enumerate}
\end{prop}

We now consider mso definable graph transductions
as a tool to specify string transductions.

There are two equally natural (and well-known)
ways of representing a string as a graph.
First, as we have seen in the Preliminaries,
for a string $w\in\Sigma^*$ of length $k$,
we may represent $w$ by the graph $\ngr(w)$ in $\GR\Sigma*$,
consisting of $k$ nodes labelled by the consecutive symbols of $w$, and
$k-1$ (unlabelled) edges representing the successor relation for the
positions of the string.
Dually, $w$ can be represented by the graph $\egr(w)$ in $\GR*\Sigma$,
consisting of $k+1$ (unlabelled) nodes, connected by $k$ edges
that form a path labelled by the symbols of $w$.
In the figure below we show $\egr(ababb)$.
Note that $\egr(\empt)$ consists of one unlabelled node.

\bigskip
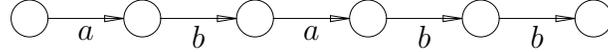
\centerline{\unitlength 0.50mm
\begin{picture}(170,20)
\input 2wgsmvbe.gtp
\end{picture}
}
\bigskip

It will turn out that the `edge graph representation' of strings is more
naturally related to two-way machines than the `node graph representation'.

\begin{defn}\leavevmode
\begin{enumerate}
\item Let $\Sigma_1, \Sigma_2$ be two alphabets,
and let
$m \subseteq\Sigma_1^*\times\Sigma_2^*$
be a string transduction.
\begin{enumerate}
\item[i.]
Its translation to graphs
$\{ ( \egr(w), \egr(z) ) \mid (w,z) \in m \}$ in
$\GR*{\Sigma_1} \times \GR*{\Sigma_2}$
is denoted by $\egr(m)$;
\item[ii.]
its translation to graphs
$\{ ( \ngr(w), \ngr(z) ) \mid (w,z) \in m \}$ in
$\GR{\Sigma_1}* \times \GR{\Sigma_2}*$
is denoted by $\ngr(m)$.
\end{enumerate}
\item
\MSOe\ denotes the family  of
all string transductions $m$
such that $\egr(m)$ belongs to \grMSO, and
\MSOn\ denotes the family  of
all string transductions $m$
such that $\ngr(m)$ belongs to \grMSO.
\end{enumerate}\unskip
\end{defn}

A transduction in \MSOe\ is called  an
{\em mso definable string transduction}, and a transduction in \MSOn\
is called a
{\em $\empt$-restricted} mso definable string transduction.
The reason for this terminology will be explained in Lemma~\ref{e=n}.

\begin{figure}[tbh]
\centerline{\unitlength 0.60mm
\begin{picture}(180,20)
\input 2n-inped.gtp
\end{picture}
}
\centerline{\unitlength 0.60mm
\begin{picture}(180,50)
\input 2n-edgr.gtp
\end{picture}
}
\caption{Edge representation for $(a^3b^2aba, a^3b^3aba)$,
	 cf.\ Example~\ref{ex-mso-edge}}\label{fig-mso-edge}
\end{figure}
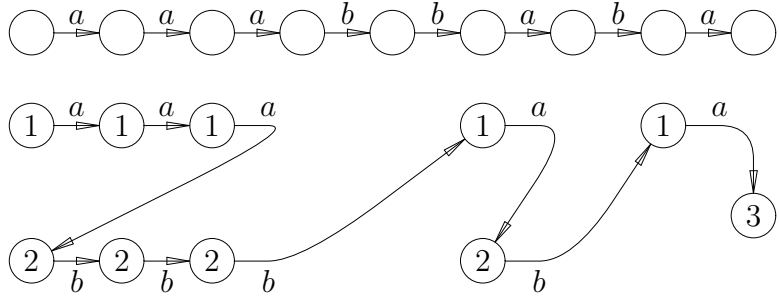

\begin{exple}\label{ex-mso-edge}
Consider the transduction  $\egr(m)$, where
$m$ is the string transduction from Example~\ref{ex1},

\[     \tradu.
\]
The formulas for the construction of the output graph
have nodes as their reference points, whereas the information
(symbols) is attached to the edges. Hence we frequently use the
formula
$\form{out}_\sigma(x) = (\exists y)\edge_\sigma(x,y)$.

As in Example~\ref{ex1a} we have an expression
$\form{fis}'_a(x,y)$
denoting the first node in the present segment of $a$'s,
this time referring to outgoing edges:
\[
y \pat x \land
	     (\forall z)( y \pat z \land z \pat x \to \form{out}_a(z) )
	  \land
	     \neg (\exists z)(\edge_a(z,y) )
\]

Similarly, we have the edge variant
$\form{next}'_a(x,y)$
by replacing  the subformulas  $\lab_a(y)$ by $\form{out}_a(y)$
in the original formula
$\form{next}_a(x,y)$.

Choosing the copy set $C=\{1,2,3\}$, and the domain formula defining
edge representations of strings,
 the transduction $\egr(m)$ is defined
by the following formulas.

\begin{Itemize}
\item[]
$\varphi^1_* = \form{out}_a(x)$
\item[]
$\varphi^2_* =
\form{out}_a(x) \land (\exists y)(x \pat y \land \form{out}_b(y) )$
\item[]
$\varphi^3_* = \neg\form{out}_a(x) \land \neg\form{out}_b(x)$, \quad
the final node of the string,
\item[]
$\varphi^{1,1}_a =  \edge_a(x,y)$
\item[]
$\varphi^{1,2}_a =  (\exists z)(\edge_a(x,z)
   \land \neg\form{out}_a(z) )
   \land \form{fis}'_a(x,y)$
\item[]
$\varphi^{1,3}_a =
	   \neg(\exists z)( \varphi^{1,1}_a(x,z)
	   \lor \varphi^{1,2}_a(x,z)  )$
\item[]
$\varphi^{2,2}_b =  \edge_a(x,y)$
\item[]
$\varphi^{2,1}_b =  (\exists z)(\edge_a(x,z) \land \neg\form{out}_a(z) )
   \land \form{next}'_a(x,y)$
\item[]
$\varphi^{2,3}_b =
	   \neg(\exists z)( \varphi^{2,1}_b(x,z)
	   \lor \varphi^{2,2}_b(x,z)  )$
\item[]
$\varphi_\sigma^{3,j} = \false$, for $j= 1,2,3$.
\end{Itemize}

The construction is illustrated in Figure~\ref{fig-mso-edge}
for $(a^3b^2aba, a^3b^3aba)\in m$. Note that we
have put the copy numbers within the nodes.
\end{exple}

The transition from one graph representation to the other
is (essentially) definable as mso graph transduction,
and will be heavily used in the sequel.
We discuss this in the next example.

\begin{exple}\label{tau}
The graph transduction {\rm ed2nd}
$= \{\; (\egr(w),\ngr(w)) \mid w\in\Sigma^*\;\}
: \GR*\Sigma \to \GR\Sigma*$
from the edge representation of a string
into its node representation is mso definable, as follows.

\begin{itemize}
\item[--]
$\varphi_\dom$ expresses that the input is a string representation,
an edge-labelled path (consisting of at least one node);
\item[--]
the copy set $C$ equals $\{1\}$;
\item[--]
$\varphi_\sigma^1 = (\exists y)( \edge_\sigma(x,y) )$
\quad,
i.e., the label $\sigma$ is moved from the edge to its source node.
None of these formulas is true for the final node of
the input graph, which means that this node is	 not	copied;
\item[--]
$\varphi_*^{1,1} = \bigvee_{\sigma\in\Sigma} \edge_\sigma(x,y)$
\quad,
i.e., edges are copied, without their labels.
\end{itemize}

\smallskip
The inverse mapping
${\rm ed2nd}^{-1} = \{\; (\ngr(w),\egr(w)) \mid w\in\Sigma^*\;\}
: \GR\Sigma* \to \GR*\Sigma$
is {\em not} mso definable:
The representation $\ngr(\empt)$ of the empty string has no nodes
that can be copied to obtain the single node of $\egr(\empt)$.

If we omit the empty string, the graph transduction
${\rm nd2ed} = \{\; (\ngr(w),$ $\egr(w)) \mid w\in\Sigma^*, w\neq\empt \;\}$
can be defined as follows.

\begin{itemize}
\item[--]
$\varphi_\dom$ again expresses that the input is a string representation,
a (non-empty) node-labelled path;
\item[--]
the copy set equals $\{1,2\}$;
\item[--]
$\varphi_*^1 = \true$, $\varphi_*^2 = \neg (\exists y)( \edge_*(x,y) )$
\quad, i.e., all nodes are copied once,
except the last one which gets two copies;
\item[--]
$\varphi_\sigma^{1,1} = \edge_*(x,y) \land \lab_\sigma(x)$
\quad, i.e., the label is moved from the node to its outgoing edge;
\item[--]
$\varphi_\sigma^{1,2} = (x=y) \land \lab_\sigma(x)$
\quad, which deals with the last edge;
\item[--]
$\varphi_\sigma^{2,1} = \varphi_\sigma^{2,2} = \false$.
\end{itemize}
\end{exple}

The above example illustrates an important technical point:
every mso graph transduction maps the empty graph to itself
(provided it belongs to the domain).
This means that, when using the node-encoding $\ngr$ for strings,
the empty string can only be mapped to itself.
As we do not want to restrict ourselves to this kind of transductions,
we have chosen to consider both variants of mso definable
string transductions.
Although $\ngr(w)$ is a slightly more direct graph representation
of the string $w$ in terms of its positions and their successor relation,
the advantage of $\egr(w)$ is that it is never empty and thus satisfies all the
usual logical laws.

\medskip
The transition from node representation to edge representation
for strings  does not influence the validity of B\"uchi's result.

\begin{prop}\label{buchi-edge}
A string language $K\subseteq \Sigma^*$ is regular iff there
is a closed formula $\varphi$ of $\MSO*\Sigma$ such that
$K=\{ w\in\Sigma^* \mid \egr(w) \models \varphi \}$.
\end{prop}

\begin{proof}
Rather direct, using B\"uchi's result
(Proposition~\ref{buchi}(2)) and Proposition~\ref{mso-closed}(2).
We consider one implication (from right to left) only.

Let the string language $K\subseteq\Sigma^*$ be defined by the closed
formula $\varphi$ of $\MSO*\Sigma$, as in the statement of the lemma
(using the edge representation).
We show that there exists a formula defining $K$ using the node
representation.
Consider the mso definable graph transduction {\rm nd2ed}
mapping $\ngr(w)$ to $\egr(w)$
for all non-empty $w\in\Sigma^*$, cf. Example~\ref{tau}.
The graph language
${\rm nd2ed}^{-1}( GL(\varphi) ) =
\{ \ngr(w) \mid  w\in\Sigma^*, w\neq\empt, \egr(w) \models \varphi \}$
is mso definable, say by an mso formula $\psi$ of $\MSO\Sigma*$.
It defines the string language
$L(\psi) = \{ w\in\Sigma^* \mid \ngr(w)\models\psi \}
  = \{ w\in\Sigma^* \mid \egr(w)\models\varphi, w\neq\empt \} =
  K-\{\empt\}$.
If $\empt\notin K$, then we are done;
  otherwise, consider $L(\psi\lor\neg(\exists x)\true)$.
\hspace*{1cm}
\end{proof}

The families \MSOn\ and \MSOe\ are equal, up to a small
technicality  involving the empty string
---a point already illustrated in Example~\ref{tau},
and in the proof of Proposition~\ref{buchi-edge}.

To prove this, we   use the following basic fact
(cf. \cite[Proposition~3.3]{msotrans}).

\begin{lem}\label{union}
Let $\tau_1$ and $\tau_2$ be mso definable graph transductions
from $\GR{\Sigma_1}{$ $\Gamma_1}$ to  $\GR{\Sigma_2}{\Gamma_2}$.
\\
If $\tau_1$ and $\tau_2$ have disjoint domains,
then also
$\tau_1 \cup \tau_2 \in \grMSO$.
\end{lem}

\begin{proof}
Consider $\tau_i$ fixed by the copy set $C_i$
and formulas $\varphi_{\dom,i}$,
$\varphi^c_{\sigma,i}$, and
$\varphi^{c_1,c_2}_{\gamma,i}$.
We may assume that $C_1$ and $C_2$ are disjoint.

The domain formula for the union is the disjunction
$\varphi_{\dom,1} \lor \varphi_{\dom,2}$;
its copy set is $C= C_1\cup C_2$.

The node formulas and the edge formulas for both transductions
are also taken together (by disjunction), 
but we ensure that they are applicable only
for the appropriate input by changing
$\varphi^c_{\sigma,i}$ to
$\varphi_{\dom,i} \land \varphi^c_{\sigma,i}$,
and similarly for the edge formulas.
We add
$\varphi^{c_1,c_2}_{\gamma}=\varphi^{c_2,c_1}_{\gamma}=\false$
for $c_1 \in C_1$, $c_2 \in C_2$,
$\gamma\in\Gamma_2$.
\hspace*{1cm}
\end{proof}

\begin{lem}\label{e=n}
Let $m\subseteq \Sigma_1^* \times \Sigma_2^*$ be a string transduction.
Then\\
$m \in \MSOn$	iff
$m \in \MSOe$ and
$(\empt,z) \in m$ implies $z=\empt$.
\end{lem}

\begin{proof}
(1) From left to right; assume $m\in \MSOn$,
i.e., $\ngr(m) \in\grMSO$.
We split $m$ into the mappings
$\hat{m} =\{ (w,z)\in m \mid  z\neq\empt \}$, and
$m_\empt =\{ (w,z)\in m \mid  z = \empt \}$.

As  $\ngr(m) \in\grMSO$,
also
$\egr(\hat{m}) = {\rm ed2nd} \cmp \ngr(m) \cmp {\rm nd2ed}$
is mso definable,
by Proposition~\ref{mso-closed}(1).

By Proposition~\ref{mso-closed}(2),
the domain of $\egr(m_\empt)$ is mso definable
as it is the inverse image of $\{\ngr(\empt)\}$
for the transduction
${\rm ed2nd} \cmp \ngr(m)$.
Now it is easily seen that $\egr(m_\empt) \in\grMSO$ using for
   $\varphi_\dom$  the formula defining the domain of $\egr(m_\empt)$,
   $C=\{1\}$,
   $\varphi^1_* = \neg(\exists y)\edge(x,y)$,
and
   $\varphi^{1,1}_\gamma = \false$.

The union $\egr(m) = \egr(\hat{m}) \cup \egr(m_\empt)$
is mso definable by Lemma~\ref{union}.
Hence, $m\in \MSOe$.
We have discussed already that the image of $\empt$ under
$m$ must be $\empt$
(provided $\empt$ belongs to the domain of $m$)
as $\ngr(\empt)$ has no nodes to copy.

\smallskip\noindent
(2) From right to left; assume $m\in \MSOe$,
i.e., $\egr(m) \in\grMSO$.

Then also $\ngr(\hat{m}) = {\rm nd2ed} \cmp \egr(m) \cmp {\rm ed2nd}$
is mso definable,
where $\hat{m} = m - \{(\empt,\empt)\}$.

We are ready when $\empt$ does not belong to the domain of $m$.
Otherwise,
as the transduction $\{(\ngr(\empt),\ngr(\empt))\}$,
mapping the empty graph to itself,
is easily seen to be mso definable,
$\ngr(m)  \in\grMSO$ follows by Lemma~\ref{union}.
\hspace*{1cm}
\end{proof}

We finally observe that, from
 Proposition~\ref{mso-closed}(1), it immediately follows
that \MSOe\ is closed under composition.
Together with the closure under composition of \DGSM\
(Proposition~\ref{prop:2dgsm-composition})
this has been a strong indication for the equality of these two families,
proved in the next section.



\section{Logic and Machines}\label{sect:logic&machine}

In this section we establish our main result,
the equivalence of
the deterministic two-way sequential machines from
Section~\ref{sect:2way},
and the mso definable string transductions
from Section~\ref{sect:mso}:
$\MSOe = \DGSM$.

The first steps towards this result were taken
already in Section~\ref{sect:2way} when we introduced
the 2gsm with mso instructions,
and showed its equivalence to the basic two-way generalized
sequential machine.

One technical notion that will be essential
to bridge the final gap between logic and machine
is modelled after Figure~\ref{fig1}
in Example~\ref{ex1}.
That figure depicts the computation of a 2gsm on a given input string.
The input string $w$ can naturally be represented by $\ngr(\tape{w})$ with
nodes corresponding to positions on the tape.
On the other hand, 
the output string $z$ is represented as $\egr(z')$
where the edges conveniently correspond to steps of the
2gsm from one position to another
(and where $z$ is obtained from $z'$ by erasing $\empt$, i.e.,
by removing the unlabelled edges).

We introduce a notation for this representation.
Let $m: \Sigma_1^* \to \Sigma_2^*$
be a string transduction.
We  use $\gr(m)$ to denote the graph transduction
$\{\; ( \ngr(\tape{w}), \egr(z)) \mid (w,z)\in m \; \}$
from $\GR{\Sigma_1\cup\{\ltape,\rtape\}}*$
to   $\GR*{\Sigma_2}$.

\begin{exple}\label{ex2}
Consider the transduction  $\gr(m)$, where
$m$ is the string transduction from Example~\ref{ex1},

\[     \tradu.
\]

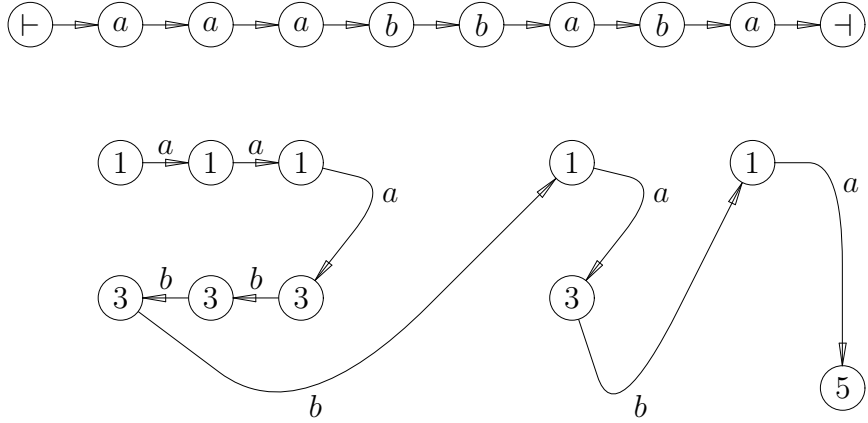
\begin{figure}[tb]
\centerline{\unitlength 0.60mm
\begin{picture}(200,20)
\input 2n-inp.gtp
\end{picture}
}
\centerline{\unitlength 0.60mm
\begin{picture}(200,80)(0,10)
\input 2n-tape.gtp
\end{picture}
}
\caption{Mso transduction $\gr(m)$ from Example~\ref{ex2}}\label{fig2}
\end{figure}

Previously we have shown that $m\in \DGSM$,
here we will demonstrate that $\gr(m)$ is an
mso definable graph transduction.

Recall the predicate
$\mbox{next}_\sigma(x,y)$  from Example~\ref{ex1a}.

For $\gr(m)$ the
domain formula specifies linear graphs of the form
$\ngr(\tape{w})$, $w\in \{a,b\}^*$,
the copy set $C$ is $\{1,3,5\}$,
and we have formulas

\begin{Itemize}
\item[]
$\varphi^1_* = \lab_a(x)$,
\item[]
$\varphi^3_* = \lab_a(x) \land (\exists y)(x\pat y \land \lab_b(y))$,
\item[]
$\varphi^5_* = \lab_\rtape(x)$,
\item[]
$\varphi_a^{1,1} = \edge_*(x,y)$,
\item[]
$\varphi_a^{1,3} = (x=y) \land \neg (\exists z)(\edge_*(x,z) \land \lab_a(z))$,
\item[]
$\varphi_a^{1,5} = \edge_*(x,y)$,
\item[]
$\varphi_b^{3,3} = \edge_*(y,x)$,
\item[]
$\varphi_b^{3,1} =
(\exists z)( \mbox{next}_b(x,z) \land \mbox{next}_a(z,y) )
\land \neg (\exists z)(\edge_*(z,x) \land \lab_a(z))$,
\quad
i.e., connect to the first $a$ of the next segment when
      we are at the first $a$ of the present segment,
\item[]
$\varphi_b^{3,5} =
    \neg (\exists z)( \varphi^{3,1}_b(x,z) \lor \varphi^{3,3}_b(x,z) ) $,
\item[]
$\varphi_\sigma^{i,j} = \false$, in all other cases.
\end{Itemize}

Note that the output of the transduction
(cf. the lower graph in Figure~\ref{fig2})
is obtained by contracting unlabelled paths in the
computation graph of the 2dgsm from Example~\ref{ex1},
Figure~\ref{fig1}.
\end{exple}

The observation from the example is generally true:
a string transduction $m$ is realized by a 2dgsm
if and only if its graph representation $\gr(m)$
is mso definable.
We prove the two implications separately.

\begin{lem}\label{lem:dgsm-mso}\label{=>tape}
Let  $m : \Sigma_1^* \to \Sigma_2^*$ be a string transduction.

\noindent
If $m\in \DGSM$,  then $\gr(m)\in \grMSO$.
\end{lem}

\begin{proof}
Let $\cM = (Q,\Sigma_1,\Sigma_2,\delta,q_\ini,q_f)$
be a 2dgsm realizing the string transduction
$m : \Sigma_1^*\to \Sigma_2^*$,
and consider a fixed input string $w=\sigma_1\cdots \sigma_n$,
$\sigma_i\in\Sigma_1$  for $i=1,\dots,n$.
Additionally we use $\sigma_0=\ltape$  and  $\sigma_{n+1} = \rtape$.

\newcommand{\cgamma}{\gamma_\cM}
We can visualize the `computation space' of $\cM$ on $w$ by constructing
a graph $\cgamma(w)$ that has as its nodes the pairs  $\paar{p,i}$,
where $p$ is a state of $\cM$, and $i\in \{0,1,\dots,n,n+1\}$ is
one of the positions of the input tape carrying $\tape{w}$.
The edges of $\cgamma(w)$ are chosen in accordance with the instruction
set $\delta$ of $\cM$:
for each instruction
   $t=(p,\sigma,\; q_1, \alpha_1,\epsilon_1,\; q_0,\alpha_0,\epsilon_0)$
in $\delta$ there is an
edge
from $\paar{p,i}$ to $\paar{q_1,i+\epsilon_1}$
if $\sigma_i$ equals $\sigma$,
and an edge
from $\paar{p,i}$ to $\paar{q_0,i+\epsilon_0}$	otherwise.
The edge is labelled by the output symbol
$\alpha_i \in \Sigma_2\cup \{\empt\}$.
In this context we will consider $\empt$ as a labelling symbol
(rather than as a string of length zero)
in order to avoid notational complications.

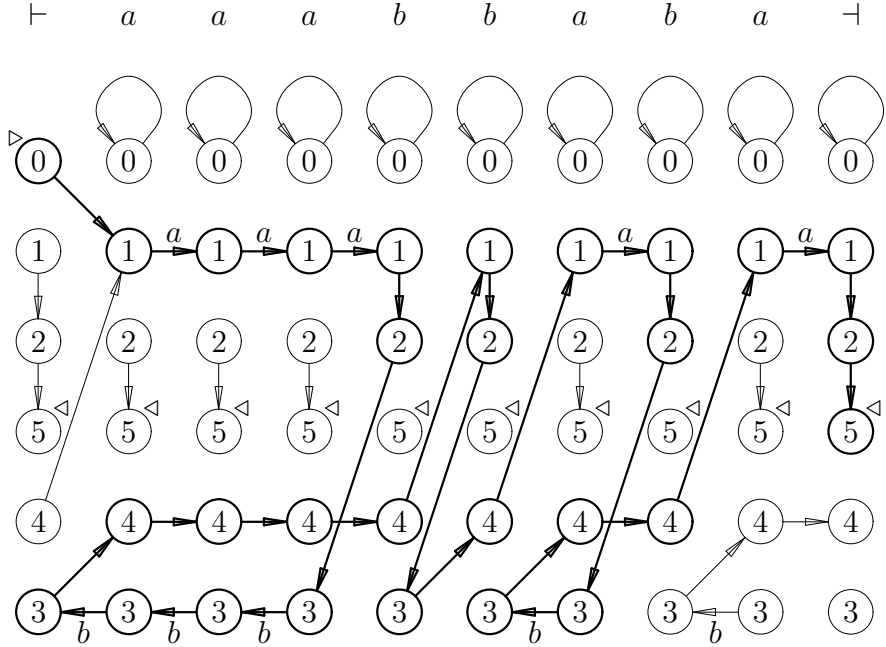
\begin{figure}[tb]
\centerline{\unitlength 0.60mm
\begin{picture}(20,140)
\input 2n-spcl.gtp
\end{picture}%
\begin{picture}(20,140)
\input 2n-spca.gtp
\end{picture}%
\begin{picture}(20,140)
\input 2n-spca.gtp
\end{picture}%
\begin{picture}(20,140)
\input 2n-spca.gtp
\end{picture}%
\begin{picture}(20,140)
\input 2n-spcb.gtp
\end{picture}%
\begin{picture}(20,140)
\input 2n-spcb.gtp
\end{picture}%
\begin{picture}(20,140)
\input 2n-spca.gtp
\end{picture}%
\begin{picture}(20,140)
\input 2n-spcb0.gtp
\end{picture}%
\begin{picture}(20,140)
\input 2n-spca0.gtp
\end{picture}%
\begin{picture}(20,140)
\input 2n-spcr.gtp
\end{picture}%
}
\caption{Computation space $\cgamma(a^3b^2aba)$ for the
2dgsm	$\cM$ in Example~\ref{ex1}%
}\label{fig5}
\end{figure}

In Figure~\ref{fig5} we illustrate the computation space for the 2dgsm
from Example~\ref{ex1} on input $a^3b^2aba$
(with output $\empt$ omitted, as usual). 
The computation on that input
is represented as a bold path   (cf. Figure~\ref{fig1}).

As $\cM$ is deterministic, every node of $\cgamma(w)$ has at most one
outgoing edge.
 The output of the computation of $\cM$ on $w$ can then
be read from $\cgamma(w)$
by starting in node $\paar{q_\ini,0}$, representing $\cM$ in its
initial configuration, and following the path along the outgoing edges.
The computation is successful if it ends in a final configuration
$\paar{q_f,k}$.
We will mark the initial and final nodes of $\cgamma(w)$ by special labels
$\ininode$ and $\finnode$, the other nodes remain unlabelled
(represented in our specification by `$\unlab$').

Note that the graph $\cgamma(w)$ does not only represent the computation of
$\cM$ on $w$ starting in the initial state and $0$-th position of the tape
(marked by $\ltape$)
but rather all possible computations that result from placing $\cM$ on an
arbitrary position of the tape, in an arbitrary state.

\medskip
We construct a series of mso graph transductions,
the composition of which maps $\ngr(\tape{w})$ to $\egr(z)$
for each $(w,z)\in m$.
As \grMSO\ is closed under
composition (Proposition~\ref{mso-closed}),
this proves the lemma.

The first graph transduction $\tau_1$
maps $\ngr(\tape{w})$ to $\cgamma(w)$.
The second graph transduction $\tau_2$ selects the path in $\cgamma(w)$
corresponding to the successful computation of $\cM$ on $w$ (if it exists)
by keeping only those nodes that are reachable from the initial
configuration and lead to a final configuration.
The last graph transduction $\tau_3$ removes edges labelled by
$\empt$ (used as a symbol representing the empty string)
while contracting paths  consisting of these edges.

\bigskip
\noindent{\em Step one: constructing $\cgamma(w)$}.
Let
$\tau_1: \GR{\Sigma_1\cup\{\ltape,\rtape\}}* \to
       \GR{\{\unlab,\ininode,\finnode\}}{$ $\Sigma_2\cup\{\empt\}}  $
be the graph transduction that constructs $\cgamma(w)$.
We follow the general description above, and formalize
$\tau_1$ as mso transduction.

The domain formula of the transduction specifies that
the graph is of the form $\ngr(\tape{w})$ for some string $w$.
The copy set equals
$C= Q$, where $Q$ is the set of states of $\cM$.
The node $\paar{q,i}$ of $\cgamma(w)$ is identified with
$u_i^q$, the $q$-copy of the node $u_i$
 of $\ngr(\tape{w})$ corresponding to the $i$-th
position of the input tape, labelled with $\sigma_i$.

\smallskip
The labels of the edges are chosen according to the instructions of $\cM$.
For $\alpha\in\Sigma_2\cup\{\empt\}$,  $p,q\in Q$,
and $\epsilon\in\{-1,0,+1\}$ let
$\step[\epsilon]_\alpha^{p,q}(x)$ be the following disjunction,
where the unspecified `dots' range over their respective components:

\[
\bigvee_{(p,\sigma,q,\alpha,\epsilon,.,.,.)\in\delta} \lab_\sigma(x) \lor
\bigvee_{\stapel{(p,\tau,.,.,.,q,\alpha,\epsilon)\in\delta}{\tau\neq \sigma}}
	 \lab_{\sigma}(x)
\]

Then,
\begin{eqnarray*}
 \varphi_\alpha^{p,q} =
 &  &	  (   \edge_*(x,y) \land \step[+1]_\alpha^{p,q}(x) )\\
 &\lor&   (	 x=y	   \land \step[0]_\alpha^{p,q}(x)	 )\\
 &\lor&   (   \edge_*(y,x) \land \step[-1]_\alpha^{p,q}(x) )\\
\end{eqnarray*}

All copies of the nodes are present,
with special labels for initial and final nodes:

\begin{Itemize}\item[]
$\varphi^q_\ininode =  \lab_\ltape(x)$,  when $q=q_\ini$,
and
$\varphi^q_\ininode = \false$,  otherwise.
\item[]
$\varphi^q_\finnode = \true$,  when $q=q_f$,
and
$\varphi^q_\finnode = \false$,  otherwise.
\item[]
$\varphi^q_\unlab =
    \neg\varphi^q_\ininode(x) \land \neg\varphi^q_\finnode(x)$.
\end{Itemize}

Note that we assume that $q_\ini \neq q_f$,
in order to avoid that both
$\varphi^q_\ininode$ and
$\varphi^q_\finnode$ are defined for the initial node.
This is the case when  $\cM$ accepts any input in its
initial state without executing instructions.
We satisfy the assumption by adding additional
instructions to a new final state.

\bigskip
\noindent{\em Step two: selecting the computation path}.
The transduction
$\tau_2: \GR{\{\unlab,\ininode,\finnode\}}{$ $\Sigma_2\cup\{\empt\}} \to
	 \GR*{\Sigma_2\cup\{\empt\}} $
removes nodes that are not on the path from the node labelled
by $\ininode$ to a node labelled by $\finnode$ (if it exists).
Nodes that are not
on such a path do not correspond to the configurations
that are part of the (successful) computation of $\cM$ on $w$.
Note that if such a path exists, then it is unique.

Recall that the predicate $\pat$
specifies the existence of a path from $x$ to $y$.
By $x\pat_\empt y$ we restrict ourselves below to a path containing
only edges with label $\empt$.

Formally,
\begin{Itemize}\item[]
$\varphi_\dom =
		(\exists x)(\exists y)[
    \lab_\ininode(x) \land \lab_\finnode(y) \land x\pat y ]$,
\item[]
$C=\{1\}$,
\item[]
$
 \varphi_*^1(x) =
		(\exists y)(\exists z)[
    \lab_\ininode(y) \land y\pat{x} \land
    \lab_\finnode(z) \land x\pat{z}	 ]
$
\item[]
and, for $\alpha\in\Sigma_2\cup \{\empt\}$,
$\varphi_\alpha^{1,1}(x,y) = \edge_\alpha(x,y)$.
\end{Itemize}

\bigskip
\noindent{\em Step three: contracting $\empt$-paths}.
The last graph transduction of three,
$\tau_3: \GR*{\Sigma_2\cup\{\empt\}} \to
	 \GR*{\Sigma_2} $
deletes all nodes that have an outgoing $\empt$-labelled edge,
and contracts each $\empt$-path to its last node.

This can be specified with the trivial copy set $C=\{1\}$,
node formula
$\varphi_*^1= \neg(\exists y)( \edge_\empt(x,y) )$,
and edge formulas
$\varphi_\alpha^{1,1}=
(\exists z)(\edge_\alpha(x,z) \land z\preceq_\empt y)$,
for $\alpha\in\Sigma_2$.
\end{proof}

Now that the 2dgsm has learned to understand the language of
monadic second-order logic, cf. Theorem~\ref{dgsm=dgsm-mso},
the converse of the previous result
has a rather straightforward proof.

\begin{lem}\label{tape=>}
Let  $m : \Sigma_1^* \to \Sigma_2^*$ be a string transduction.

\noindent
If $\gr(m)\in \grMSO$, then
  $m\in \DGSM$.
\end{lem}

\begin{proof}
Starting with the mso transduction
$\gr(m) : \GR{\Sigma_1 \cup\{\ltape,\rtape\}}* \to \GR*{\Sigma_2}$
we build a  2dgsm-mso $\cM$ for $m$ that closely follows the
mso specification of  $\gr(m)$.

Assume $\gr(m)$ is specified by
domain formula $\varphi_\dom$,
copy set $C$,
node formulas $\varphi^{c}_*$, $c\in C$, and
edge formulas $\varphi_\sigma^{c_1,c_2}$,
		      $c_1,c_2\in C$, $\sigma\in\Sigma_2$.
The state set of $\cM$ is (in principle) equal to the copy set $C$:
when
$  \varphi_\sigma^{c_1,c_2}(u,v)$
is true for a pair $u,v$
of nodes, then $\cM$, visiting the position corresponding to $u$
of the input tape in state $c_1$,
   may move to the position corresponding to $v$ changing to state $c_2$,
while writing $\sigma$ to the output tape.

Note that, for each input graph $g$,
$\gr(m)(g)$ defines a graph representation of a string,
hence at most one of these formulas defines an edge in a given position (node)
and a given state (copy).
However, in general the formula
$\varphi_\sigma^{c_1,c_2}$ is only functional as far as graphs $g$
satisfying the domain formula $\varphi_\dom$ are concerned,
and for these graphs only when restricted to nodes for which the
respective $c_1$ and $c_2$ copies are defined.
Since our formal definition of 2dgsm-mso demands functional moves,
 we consider the formulas
$\psi_\sigma^{c_1,c_2}(x,y) =
\varphi_\sigma^{c_1,c_2}(x,y) \land
   \varphi^{c_1}_*(x) \land
   \varphi^{c_2}_*(y) \land \varphi_\dom$.

The instructions of $\cM$ are of the form
\[ (  c_1, (\exists y)( \psi_\sigma^{c_1,c_2}(x,y) ), \;
    c_2, \sigma, \psi_\sigma^{c_1,c_2}(x,y) \;)
\]
--
but this is 5-tuple notation,
and has to be replaced by 8-tuples where
for a fixed state $c_1$
each of the alternatives
$(c_2, \sigma) \in C\times \Sigma_2$ has to be tested consecutively,
as explained in the paragraph about 2gsm in Section~\ref{sect:2way}
(using additional states).

If none of the edge formulas gives a positive result, the
present node has no successor, which indicates the last position of the 
output string.
In that case, the series of consecutive tests ends up in the
final state $q_f$.

Initially $\cM$ has to find the unique node of the output graph
that has no incoming edges.
We solve this by adding the new initial state $q_\ini$
from which this node is found by
testing all possibilities,
but again in a consecutive fashion,
for $c_2 \in C$:
\[
(  q_\ini,
    (\exists y)[ \varphi^{c_2}_*(y) \land
    \neg\inco^{c_2}(y) ], \;
    c_2, \empt,
    \varphi^{c_2}_*(y) \land
    \neg\inco^{c_2}(y) \;)
\]
where $\inco^{c_2}(y)$ abbreviates
$(\exists z)\bigvee_{{c_1\in C},{\sigma\in \Sigma_2}}
(\psi_\sigma^{c_1,c_2}(z,y) )$.
\end{proof}

\begin{lem}\label{lem:id}
Let $\Sigma$ be an alphabet.
The transduction $\gr(id)
 : \GR{\Sigma \cup\{\ltape,\rtape\}}* \to \GR*{\Sigma}$
mapping $\ngr(\tape{w})$ to $\egr(w)$
is an element of $\grMSO$,
as is its inverse $\gr(id)^{-1}$.
\end{lem}

\begin{proof}
The identity on $\Sigma^*$ is easily performed by an 2dgsm.
Hence $\gr(id) \in \grMSO$, by Lemma~\ref{lem:dgsm-mso}.

As for the inverse  $\gr(id)^{-1}$,
note that mapping $\egr(w)$ to $\egr(\tape{w})$ is mso definable
because $\egr(w)$ has at least one node, which may be copied
to provide the additional nodes that are connected by edges
labelled by $\ltape$ and $\rtape$ to the original graph.
We now compose this mapping by {\rm ed2nd},
which is mso definable by Example~\ref{tau}.
\end{proof}

We complete the section by deriving the
equivalence between the mso definable string transductions and
the deterministic two-way finite state transductions,
uniting logic and machines.

\begin{thm}\label{thm-logic&machine}
$\MSOe = \DGSM$.
\end{thm}

\begin{proof}
By 
our previous lemma,
the transduction $\gr(id)$ from $\ngr(\tape{w})$ to $\egr(w)$,
for $w\in\Sigma_1^*$, is an element of $\grMSO$,
as is its inverse $\gr(id)^{-1}$.
By the equalities $\gr(m) = \gr(id) \cmp \egr(m)$, and
$\egr(m) = \gr(id)^{-1} \cmp \gr(m)$, and the closure of \grMSO\ under
composition (Proposition~\ref{mso-closed}), we have
$m \in \MSOe$ iff (by definition)
$\egr(m) \in \grMSO$ iff
$\gr(m)\in \grMSO$.

The result now follows from Lemmas~\ref{=>tape} and \ref{tape=>}
demonstrating $\gr(m) \in \grMSO$ iff $m \in \DGSM$.
\end{proof}

As an immediate consequence of this result and Lemma~\ref{e=n}
we obtain the equivalence between the corresponding $\empt$-restricted
transductions.

We use \DGSMl\ to denote those relations $m$ in \DGSM\
that satisfy
$(\empt, z) \in m$ implies $z=\empt$,
cf.
Lemma~\ref{e=n}.

\begin{cor}
$\MSOn = \DGSMl$.
\end{cor}


\section{Nondeterminism}\label{sect:nondet}

In this section we define the nondeterministic mso definable
graph transductions, and their derived string relatives.
We observe that nondeterministic mso transductions are
related to the deterministic mso transductions via
relabelling of the input.

\medskip
A nondeterministic variant of
 mso definable transductions is considered in \cite{msoV,msotrans}.
All the formulas of the deterministic version
may now have additional free node-set variables
$X_1,\dots,X_k$,
called	   `parameters',
the same   for each of the formulas.
For each valuation of the parameters (by sets of nodes of the input graph)
that satisfies the domain formula,
the other formulas define the output graph as before.
Hence each valuation may lead to a different output graph
 for the given input graph: nondeterminism.

\medskip
More formally,
a {\em nondeterministic mso definable (graph) transduction}
$\tau \subseteq \GR{\Sigma_1}{\Gamma_1} \times \GR{\Sigma_2}{\Gamma_2}$
is specified by

\begin{Itemize}
\item[--]
       a set of {\em parameters} $X_1,\dots,X_k$, $k\ge 0$,
\item[--]
       a {\em domain formula} $\varphi_{\dom}(X_1,\dots,X_k)$,
\item[--]
       a finite {\em copy set} $C$,
\item[--]
      {\em node formulas} $\varphi_{\sigma}^{c}(x,X_1,\dots,X_k)$
      for $\sigma\in \Sigma_2$, $c\in C$, and
\item[--]
      {\em edge formulas} $\varphi_{\gamma}^{c_1,c_2}(x,y,X_1,\dots,X_k)$
      for $\gamma\in\Gamma_2$, $c_1,c_2\in C$,
\end{Itemize}

where all formulas are in $\MSO{\Sigma_1}{\Gamma_1}$.
\medskip

Recall from Section~\ref{preliminaries} that an input graph together with a
valuation of the parameters can be represented by a $\Xi$-valuated
graph $g$
which has node labels in $\Sigma_1\times\pow{\Xi}$
(where $\Xi= \{ X_1,\dots,X_k \}$)
such that $g\restrict \Sigma_1$ is the input graph, and $\nu_g$ is the
valuation.
By definition, $g\in GL(\varphi_\dom)$ iff
$g\restrict\Sigma_1, \nu_g \models \varphi_\dom(X_1,\dots,X_k)$.

For each $g\in GL(\varphi_\dom)$
we define the graph $\hat\tau(g)$ similar to $\tau(g)$ in
Definition~\ref{def-mso}.
The nodes of $\hat\tau(g)$
are defined using
$g\restrict \Sigma_1 \models \varphi^{c}_\sigma(u,U_1,\dots,U_k)$,
where $U_i = \nu_g(X_i)$,
rather than
$g \models \varphi^c_\sigma(u)$,
and similarly for the edges and node labelling of $\hat\tau(g)$.
The transduction $\tau$ is then defined as follows:
$\tau= \{\; (g\restrict \Sigma_1, \hat\tau(g))
       \mid g\in GL( \varphi_\dom) \;\}$.

\begin{exple}\label{not-gsm}
Let $m \subseteq \{ a\}^* \times \{a,b,\#\}^*$
be the relation
\[
\{\; (a^n, w\# w) \mid n\ge 0, w\in\{a,b\}^*, |w|=n \;\}.
\]

The relation $\egr(m)$ can be realized by a nondeterministic
mso definable transduction,
with parameters $X_a$ and $X_b$.
The nodes of the input graph are copied twice, and
the parameters determine whether
the outgoing edge of a node in the input is copied as
$a$-edge or $b$-edge, respectively.

The components of the transduction are as follows.
The copy set equals $C = \{1,2\}$,
the domain formula $\varphi_\dom(X_a,X_b)$
expresses that the input graph is a string
representation, and additionally that the sets $X_a$ and $X_b$
form a partition of its nodes.

All input nodes are copied twice:
$\varphi^1_*(x,X_a,X_b)
= \varphi^2_*(x,X_a,X_b)  = \true$.

The edge labels are changed
according to the sets $X_a$ and $X_b$, additionally
the last node of the first copy is connected to the
first node of the second copy by an $\#$-edge:

\begin{Itemize}
\item[]
$\varphi^{1,1}_\sigma(x,y,X_a,X_b)
= \varphi^{2,2}_\sigma(x,y,X_a,X_b)
	  = \edge_a(x,y) \land x\in X_\sigma$,	\\
	  \phantom{XX} \hfill for $\sigma=a,b$,
\item[]
$\varphi^{1,2}_\#(x,y,X_a,X_b) =
 \neg( \exists z) \edge_a(x,z)
 \land
 \neg( \exists z) \edge_a(z,y)
$,
\item[]
$\varphi^{i,j}_\sigma(x,y,X_a,X_b) = \false$,
for all other combinations $i,j,\sigma$.
\end{Itemize}
Mapping $aaa$ to $abb\#abb$ can be realized by taking
the valuation
$\nu(X_a)=\{1\}$, $\nu(X_b)=\{2,3,4\}$.

%
\medskip
Note that this example can be changed such that it uses only one
parameter, as the sets represented by the parameters are complementary.
\end{exple}

We use \grNMSO, \NMSOn, and \NMSOe\
to denote the nondeterministic counterparts of the
families \grMSO, \MSOn, and \MSOe, respectively.
The family of (nondeterministic) 2gsm transductions
is denoted by \NGSM.

\bigskip
Unlike the deterministic case,
the power of the nondeterministic 2gsm
is incomparable to that of the
nondeterministic mso definable string transduction.
First,
because the number of parameter valuations is finite, every 
nondeterministic mso transduction
is finitary.
This is not true for the 2gsm,
which can realize the (non-finitary) transduction
$\{\; (a^n,a^{mn}) \mid m,n \ge 1 \;\}$,
by nondeterministically choosing the number $m$ of copies made of the input.

On the other hand,
the nondeterministic mso  transduction of the previous example
cannot be realized by a 2gsm.

\begin{lem}\label{lem:not-gsm}
Let $m \subseteq \{a\}^* \times \{a,b,\#\}^*$
be the relation
$
\{\; (a^n, w\# w) \mid n\ge 0, w\in\{a,b\}^*, |w|=n \;\}
$.
Then  $m \notin \NGSM$.
\end{lem}

\begin{proof}
Assume $m$ is realized by a (nondeterministic) 2gsm $\cM$
with $k$ states.  Choose  $n$ such that $2^n > k\cdot (n+2)$.
Consider the behaviour of $\cM$ on input $a^n$.
The input tape, containing $\tape{a^n}$, has $n+2$ positions.
Hence, $\cM$ has $k\cdot (n+2)$ configurations on this input.
Consider the configuration assumed by $\cM$ when it has just written the
symbol $\#$ on its output tape.
As there are $2^n$ possible output strings $w\#w$ for $a^n$,
there exist two strings $w_1$ and $w_2$ for which this configuration
is the same.
This means that we can switch the computation of $(a^n, w_1\#w_1)$
halfway to the computation of $(a^n, w_2\#w_2)$ obtaining a
computation for $(a^n, w_1\#w_2)$ with $w_1 \neq w_2$,
which is not an element of $m$.
\end{proof}

It is not difficult to see that the relation $m$ from the lemma,
can be realized by the composition of two 2gsm's,
the first nondeterministically mapping $a^n$ to a string $w\in \{a,b\}^*$
with $|w|=n$,
the second (deterministically) doubling its input $w$ to $w\# w$.
This shows that \NGSM\ is not closed under composition,
as proved in  \cite{kiel} for the corresponding families of output languages.
In fact, the families $\NGSM^k$ of compositions of $k$ 2gsm transductions
form a strict hierarchy, as proved in \cite{gre78c,eng82,iterated}
(again for the corresponding families of output languages).

However, the nondeterministic mso transductions
are closed under composition
\cite[Prop.~5.5.6]{cou97}.

\begin{prop}\label{grNMSO-closed}
\grNMSO,
and consequently \NMSOe\ and \NMSOn\,
are
   closed under composition.
\end{prop}

By \grREL\ we denote the family of (nondeterministic) node
relabellings for graphs. A relation in $\GR{\Sigma_1}\Gamma \times
\GR{\Sigma_2}\Gamma$
is a {\em node relabelling} if there exists a relation
$R\subseteq\Sigma_1\times\Sigma_2$
such that the images of a graph $g$ are exactly those graphs that can
be obtained from $g$ by replacing every occurrence of a node label
$\sigma$ by an element of $R(\sigma)$,
leaving edges and their labels unchanged.

We use \REL\ to denote the family of (nondeterministic)
{\em string relabellings},
related to \grREL\ through the mapping \ngr.

\medskip
We observe the following elementary relationship between deterministic and
nondeterministic mso definable graph transductions.

\begin{thm}\label{thm:grNMSO-grMSO}
$\grNMSO = \grREL \cmp \grMSO$.
\end{thm}

\begin{proof}
The proof of the first inclusion
 $\grNMSO \subseteq \grREL \cmp \grMSO$
is implicit in our  definition of \grNMSO.
The nondeterminism of an mso transduction $\tau$ with parameters
$X_1,\dots,X_k$
can be `pre-processed' by a relabelling $\rho$ that
maps each node label $\sigma\in\Sigma_1$ nondeterministically to
a symbol $(\sigma,f)\in\Sigma_1\times\pow{\Xi}$, where $\Xi=\{X_1,\dots,X_k\}$.
The valuation of $X_i$ has now become a part of the labelling,
and we change
the domain formula
      $\varphi_\dom(X_1,\dots,X_k)$,
the node formulas
       $\varphi_{\sigma}^{c}(x,X_1,\dots,X_k)$,
and the edge formulas
    $\varphi_{\gamma}^{c_1,c_2}(x,y,X_1,\dots,X_k)$
that specify the mso transduction accordingly.
Each atomic subformula $y\in X_i$ in such a formula is replaced
by the disjunction
$\bigvee_{f(X_i)=1,\sigma\in\Sigma_1} \lab_{(\sigma,f)}(y)$,
and each atomic subformula $\lab_\sigma(y)$ is replaced by
$\bigvee_{f:\Xi\to\{0,1\}} \lab_{(\sigma,f)}(y)$.
In this way we obtain `deterministic' equivalents
	$\hat{\varphi}_\dom$,
       $\hat{\varphi}_{\sigma}^{c}(x)$,
    $\hat{\varphi}_{\gamma}^{c_1,c_2}(x,y)$
for mso transduction $\hat{\tau}$.
We now have $\tau = \rho \cmp \hat\tau$
which follows by observing that for a graph
$g \in \GR{\Sigma_1\times \{0,1\}^\Xi}*$,
$g \models \hat{\varphi}_\dom$
if and only if
$g\restrict \Sigma_1, \nu_g \models \varphi_\dom(X_1,\dots,X_k)$,
and similarly for the other formulas.

\medskip
For the converse inclusion
$\grNMSO \supseteq \grREL \cmp \grMSO$,
it suffices to note that each nondeterministic node relabelling is
a nondeterministic mso definable graph transduction.
The inclusion then follows from the closure of $\grNMSO$ under composition,
Proposition~\ref{grNMSO-closed}.

Let
$R\subseteq \Sigma_1 \times \Sigma_2$
define a graph node relabelling.
We formalize it as mso graph transduction
from $\GR{\Sigma_1}\Gamma$ to $\GR{\Sigma_2}\Gamma$
by choosing parameters $X_\tau$, $\tau\in \Sigma_2$,
with the intended meaning that a node belonging to
$X_\tau$ will be relabelled into $\tau$.

The domain formula $\varphi_\dom$ expresses that the $X_\tau$
form an `admissable' parameter set by demanding
each node  to be in exactly one of the $X_\tau$,
and additionally,
if a node has label $\sigma$,
then $X_\tau$ containing this node satisfies $\tau\in R(\sigma)$:

\[
(\forall x)
  \bigvee_{\tau\in \Sigma_2}( x\in X_\tau \land
       \bigwedge_{\tau'\neq \tau} x \notin X_\tau
			    )
\land
(\forall x)
   \bigwedge_{\sigma\in \Sigma_1}(
	 \lab_\sigma(x) \to \bigvee_{\tau\in R(\sigma)} x\in X_\tau
				 )
\]

Each node is copied once, relabelled according to $X_\tau$:
\begin{Itemize}
\item[]
$C=\{ 1 \}$,
\item[]
$\varphi_\tau^1 = x \in X_\tau$, $\tau\in \Sigma_2$,
\item[]
$\varphi_\gamma^{1,1} = \edge_\gamma(x,y)$, $\gamma \in \Gamma$.
\end{Itemize}
\end{proof}

As we have observed,
any string relabelling can be `lifted' to a graph node relabelling
using the graph interpretation $\ngr$ of strings.
By restricting the previous result to those graph transductions that result
from strings,
we obtain a result for mso definable string transductions
in the node interpretation.

\begin{cor}
$\NMSOn = \REL \cmp \MSOn$.
\end{cor}

In addition to \REL, we
   need \MREL\ denoting the family of
{\em marked string relabellings},
that map a string $w$ first to the `marked version'
$\tape{w}$,
and then apply a string relabelling.

\begin{thm}
$\NMSOe = \MREL \cmp \MSOe$.
\end{thm}

\begin{proof}
First, the  inclusion from left to right.
Let $m \in \NMSOe$, i.e.,  $\egr(m) \in \grNMSO$.

Consider the string transduction
$m' = \{\; (\tape{w},z) \mid (w,z)\in m \;\}$.
Then $m'$ is an element of $\NMSOn$,
as $\ngr(m')$ equals the composition
$\gr(id)\cmp \egr(m) \cmp \mbox{\rm ed2nd}$
of (nondeterministic) mso definable graph transductions,
where $\gr(id)$ is the mapping from
$\ngr(\tape{w})$ to $\egr(w)$,
cf. Lemma~\ref{lem:id}.

By the corollary   above,
and Lemma~\ref{e=n},
$m' \in \REL \cmp \MSOn \subseteq  \REL \cmp \MSOe$.
Consequently, as $m$ equals the `marking' from $w$ to $\tape{w}$
followed by $m'$,
$m \in	\MREL \cmp \MSOe$.

\medskip
For the reverse inclusion,
$\NMSOe \supseteq \MREL \cmp \MSOe$,
note that every marked relabelling can be decomposed
into a marking and a relabelling,
each of which we will show to be a (nondeterministic) mso transduction.
The inclusion then follows from
the closure of \NMSOe\ under composition.

\smallskip
The marking mapping $w$ to $\tape{w}$ is easily seen to be an element
of \MSOe,
either by direct construction,
or by constructing a 2dgsm for that task, and applying
Theorem~\ref{thm-logic&machine}.

\smallskip
Finally,
to show that $\REL \subseteq \NMSOe$ one closely follows the
argumentation in the proof of
$\grREL \subseteq \grNMSO$, Theorem~\ref{thm:grNMSO-grMSO}.
As we relabel edges, rather than nodes, in the
representation $\egr(w)$ of a string $w$,
but still have parameters ranging over nodes,
we use the parameters for the source node of an edge to determine the new
label of its outgoing edge (cf. Example~\ref{not-gsm}):
$\varphi_\dom$ is as before, but we now have
$\varphi_*^1	    = \true$, and
$\varphi_\tau^{1,1} = \edge(x,y) \land (x\in X_\tau)$.
\end{proof}

For completeness we
note that the above result cannot be strengthened to
$\NMSOe = \REL \cmp \MSOe$,
as the relations on the right side are functional for the empty string $\empt$.
This is not necessarily true for \NMSOe.

\begin{exple}
The string transduction $\{ (\empt,a),	(\empt,b) \}$
in $a^* \times \{a,b\}^*$ is realized by the following
nondeterministic mso transduction, in the edge representation.
The single parameter $X$ determines whether $\empt$ is mapped
to $a$ or to $b$.
Let 
\begin{Itemize}
\item[]
$\varphi_\dom = (\exists x)(\; (\forall y)(y=x) \land\neg \edge_a(x,x) \;)$,
\item[]
$C = \{1,2\}$,
\item[]
$\varphi^1_* = \varphi^2_* = \true$,
\item[]
$\varphi^{1,1}_\sigma =
 \varphi^{2,1}_\sigma = \varphi^{2,2}_\sigma = \false$,
for $\sigma \in \{a,b\}$,
and
\item[]
$\varphi^{1,2}_a  = x\in X$,
\item[]
$\varphi^{1,2}_b  = \neg(x\in X)$.
\end{Itemize}
\end{exple}

Combining the previous two results
(that relate the nondeterministic and deterministic
mso transductions)
with the equalities between deterministic mso transductions
and deterministic gsm mappings
of Theorem~\ref{thm-logic&machine},
we directly obtain the following result.

\begin{thm}\label{NMSO=REL;DGSM}%
\leavevmode\\
$\NMSOe = \MREL \cmp \DGSM$
\hspace{2ex}and\hspace{2ex}
$\NMSOn = \REL \cmp \DGSMl$.
\end{thm}


\section{Finite Visit Machines}\label{finite-visit}

Rajlich
\cite{rajlich}
observes that 2gsm are more powerful than 2dgsm
(as generative devices, by considering their output languages,
i.e., the ranges of the transductions).
He demonstrates that this is mainly due to the ability
of the 2gsm to visit each of the positions of its input
an unbounded number of times.

Motivated by this result,
we consider transducers that have a fixed bound on the
number of times they visit each of their input positions
--we call this the finite visit property--
and relate these to the (nondeterministic) mso transductions.

We show that the nondeterministic mso definable string transductions are
exactly those transductions that are realized by the
composition of two 2gsm with the finite visit property.
Note that one direction of this result follows from
Theorem~\ref{NMSO=REL;DGSM}.

Moreover, we characterize the nondeterministic mso definable 
string transductions as those compositions of 2gsm's that
realize finitary transductions, i.e.,
transductions that define a finite number of images for
every input string.

A more direct characterization can be obtained by considering
2gsm that are allowed to rewrite the symbols on their
input tape (but with the finite visit property).
These machines exactly match the mso
definable string transductions,
both in the deterministic case
and the nondeterministic case.

The finite visit property was studied in, e.g.,
\cite{hennie,rajlich,gre78a,gre78b,gre78c,ers80,eng82}.

\subsection{Finite visit two-way generalized sequential machines}

A computation of a 2gsm is called $k$-{\em visiting} if each of the
positions of the input tape is visited at most $k$ times.
The 2gsm $\cM$ is called {\em finite visit},
if there is a constant $k$ such that,
for each pair $(w,z)$ in the transduction realized by $\cM$,
there exists a $k$-visiting computation for $(w,z)$.
The family of string transductions realized by finite
visit nondeterministic 2gsm is denoted by \NGSMf.

Note that our definition is rather weak,   as the machine may have
many computations that are not $k$-visiting, either without
any chance of reaching the final state,
or with loops in the computation that produce no output.

If a deterministic 2gsm visits a position of the input tape twice in
the same state, then the computation will enter an infinite loop that
will not reach the final state. This implies 
the well-known fact that every deterministic
2gsm is finite visit, where we choose for $k$ the number of states of
the machine.
A similar argument  enables us to prove the following characterization
of finite visit transductions
in terms of transductions that map each input string into a finite
number of output strings.

\begin{lem}\label{finitary}
Let $m$ be a string transduction. Then
\\
$m \in \NGSMf$ iff
     $m \in \NGSM$ and $m$ is finitary.
\end{lem}

\begin{proof}
Clearly, the length of the output of a $k$-visiting computation on input $w$
is at most $k$ times the length of $\tape{w}$. Hence the implication from
left to right.

As for the other implication, assume that the finitary	   transduction
$m$ is realized by a 2gsm $\cM$.
If during a (successful) computation for $(w,z)\in m$,
$\cM$ visits the same position
twice in the same state, then it did not write symbols to the output
in the meantime, because otherwise
$\cM$ has infinitely many output strings for the present input, as
an easy pumping argument shows.
Hence we may omit this excursion from the computation.
Consequently, there is a computation of $\cM$ for $(w,z)$ that does not
visit each of the tape positions more than $k$ times, where $k$ is the
number of states of $\cM$.
Hence $\cM$ {\em itself} is finite visit.
\end{proof}

It is well known
(see, e.g., \cite{fis69,chytil,gre78a,gre78b,ahul70})
that the computation of a finite visit 2gsm on an input tape
can be coded as a string of `visiting sequences'
(strongly related to `crossing sequences',
cf. \cite{rab63,hennie,houl-book,birget}).
We recall how this can be done, without going into details.

We consider several types of visits during a computation,
differing  in the direction
($-1$, $0$, or $+1$)
of the steps
taken by the machine just before and just after the visit.
Additionally, a visit may be either the first or the last visit of the
computation.

Given a computation of a 2gsm,
the {\em visiting sequence} of a position of the input tape is the
sequence that starts with the symbol $\sigma$
on the tape, followed by the
consecutive visits of the machine to that position.
Each of the visits is given as a 4-tuple
$(\mineps,p,\pluseps,\alpha)$
consisting of
the direction $\mineps$ of the move before the visit,
the state $p$ during the visit,
the direction $\pluseps$ of the move after the visit,
and the string $\alpha$ written to the output during that move.
For the first visit we take $\mineps=*$,
for the last visit we take $\pluseps=*$.

We illustrate this notion with an example.

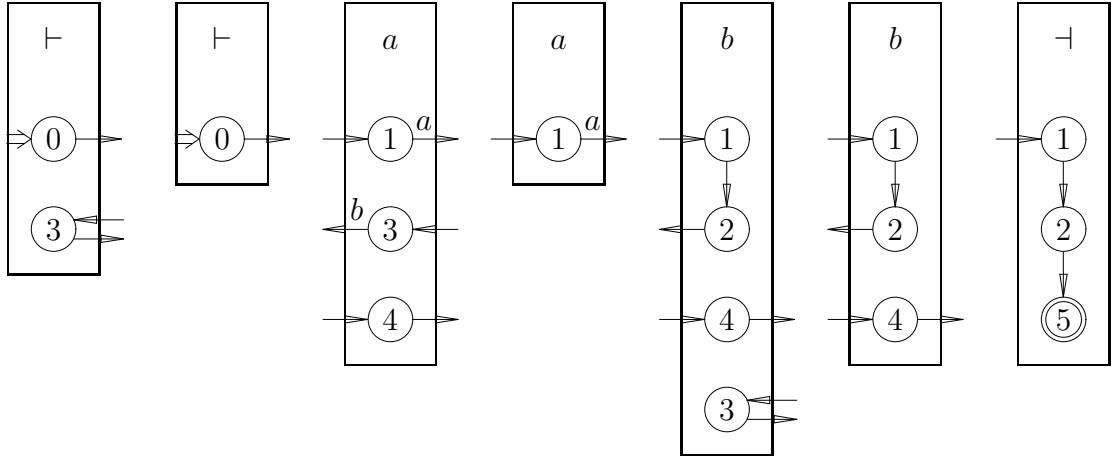
\begin{figure}
\centerline{\unitlength 0.60mm
\begin{picture}(35,120)
\input 2n-visl.gtp
\end{picture}
\begin{picture}(35,120)
\input 2n-visl0.gtp
\end{picture}
\begin{picture}(35,120)
\input 2n-visa.gtp
\end{picture}
\begin{picture}(35,120)
\input 2n-visa0.gtp
\end{picture}
\begin{picture}(35,120)
\input 2n-visb.gtp
\end{picture}
\begin{picture}(35,120)
\input 2n-visb0.gtp
\end{picture}
\begin{picture}(35,120)
\input 2n-visr.gtp
\end{picture}
}
\caption{Visiting sequences for Example~\ref{ex1},
cf. Example~\ref{ex:sequences}.}\label{fig-vis}
\end{figure}

\begin{exple}\label{ex:sequences}
Consider the 2dgsm from Example~\ref{ex1}.
Each of the visiting sequences during a successful computation
is one of the following.

\begin{Itemize}
\item[]
$\paar{\; \ltape, \; (*,0,+1,\empt), (-1,3,+1,\empt)   \; }$
\item[]
$\paar{\; \ltape, \; (*,0,+1,\empt)   \; }$
\item[]
$\paar{\; a, \; (+1,1,+1,a), (-1,3,-1,b), (+1,4,+1,\empt) \; }$
\item[]
$\paar{\; a, \; (+1,1,+1,a) \; }$
\item[]
$\paar{\; b, \; (+1,1,0,\empt), (0,2,-1,\empt), (+1,4,+1,\empt),
(-1,3,+1,\empt)\; }$
\item[]
$\paar{\; b, \; (+1,1,0,\empt), (0,2,-1,\empt), (+1,4,+1,\empt) \; }$
\item[]
$\paar{\; \rtape, \; (+1,1,0,\empt), (0,2,0,\empt), (0,5,*,\empt)\; }$
\end{Itemize}
These visiting sequences are depicted in a suitable graphical manner in
Figure~\ref{fig-vis},
cf.~Figure~\ref{fig1}.
\end{exple}

Each visiting sequence must satisfy some syntactical constraints.

First, the directions of the visits are `alternating'.
This means that
the first visit enters from the left ($\mineps=+1$,
with the exception for $\sigma=\ltape$ which  starts in the initial
state with $\mineps=*$);
then,
 if the move after the $i$-th visit equals
$\pluseps=-1,0,+1$, then the move prior to the $i\!+\!1$-st visit to the
same position must equal $\mineps'=+1,0,-1$, respectively.
Only the last visit of a sequence can have $\pluseps=*$,
in case the state is final,
signalling the end of a computation.

Secondly, the direction $\pluseps$
of the move after the visit, and the string
$\alpha$
written to the output,
must correspond to an instruction of the machine for the
given input symbol $\sigma$ and the given state $p$.
Additionally,
when $\pluseps=0$,
the new state given by the instruction must match the next visit of the
visiting sequence.

Clearly, also neighbouring visiting sequences for a given computation
must satisfy several constraints.
If a visiting sequence has $k$ `crossings' to the right,
either outgoing visits $(\mineps,p,+1,\alpha)$
or incoming visits $(-1,p,\pluseps,\alpha)$
--they alternate--
then the visiting sequence to the right has
exactly $k$ matching crossings to the left,
matching both in direction
(which implicitly follows from the restrictions on single
visiting sequences above)
and in state  change for  the machine.
Note that a visit $(-1,p,+1,\alpha)$ represents two crossings.

Finally, the first visiting sequence of a computation should start
with a visit $(*,q_\ini,\pluseps,\alpha)$,
and exactly one visiting sequence should end with a visit 
$(\mineps,q_f,*,\alpha)$.

\smallskip
When we bound the number of visits to each position,
the visiting sequences
come from a finite set, and we can interprete these sequences as symbols
from a finite alphabet.
Each $k$-visiting computation is specified by a string over this alphabet,
and we will call these strings $k$-{\em tracks}.
(E.g., the track
in Figure~\ref{fig-vis2} specifies
the computation of the 2dgsm of Example~\ref{ex1} on input $a^3b^2aba$,
cf. Figure~\ref{fig1}).
It should be obvious from the above remarks that
the language of such specifications is regular
(see, e.g., Lemma~2.2 of \cite{gre78a},
or Lemma~1 of \cite{chytil}).
For instance, it
is the heart of the  proof in \cite[Theorem~2.5]{houl-book} of the result
that two-way finite state automata
are equivalent to their one-way counterparts \cite{2to1a,2to1b}.

\begin{figure}
\centerline{\unitlength 0.50mm
\begin{picture}(40,120)
\input 2n-visl.gtp
\dottedline{1.3}(25,70)(35,70)
\dottedline{1.3}(25.5,52.1)(35,50)
\dottedline{1.3}(25.5,47.9)(35,30)
\end{picture}%
\begin{picture}(40,120)
\input 2n-visa.gtp
\put(25,105){\makebox(0,0)[cc]{$\times$3}}
\dottedline{1.3}(25,70)(31,70)
\dottedline{1.3}(25,50)(31,50)
\dottedline{1.3}(25,30)(31,30)
\dottedline{1.3}(35,70)(41,70)
\dottedline{1.3}(35,50)(41,50)
\dottedline{1.3}(35,30)(41,30)
\end{picture}%
\begin{picture}(10,120)
\end{picture}%
\begin{picture}(40,120)
\input 2n-visb.gtp
\dottedline{1}(25,30)(35,70)
\dottedline{1}(25.5,12.1)(35,50)
\dottedline{1}(25.5,07.9)(35,30)
\end{picture}%
\begin{picture}(40,120)
\input 2n-visb.gtp
\dottedline{1}(25,30)(35,70)
\dottedline{1}(25.5,12.1)(35,50)
\dottedline{1}(25.5,07.9)(35,30)
\end{picture}%
\begin{picture}(40,120)
\input 2n-visa.gtp
\dottedline{1.3}(25,70)(35,70)
\dottedline{1.3}(25,50)(35,50)
\dottedline{1.3}(25,30)(35,30)
\end{picture}%
\begin{picture}(40,120)
\input 2n-visb0.gtp
\dottedline{1}(25,30)(35,70)
\end{picture}%
\begin{picture}(40,120)
\input 2n-visa0.gtp
\dottedline{1.3}(25,70)(35,70)
\end{picture}%
\begin{picture}(40,120)
\input 2n-visr.gtp
\end{picture}%
}
\caption{Track
	 for $\tape{a^3b^2aba}$,
	 Example~\ref{ex1}.}\label{fig-vis2}
\end{figure}

\begin{prop}\label{regular-visiting}
Let $\cM$ be a 2gsm, and let $k$ be a constant.
The $k$-tracks
for successful $k$-visiting
computations of $\cM$ form a regular language.
\end{prop}

From this result, using standard techniques
(see e.g., \cite[Lemma~1]{chytil})
we obtain the following decomposition of finite visit
nondeterministic 2gsm transductions.
Note that this decomposition already features in
Theorem~\ref{NMSO=REL;DGSM}
as characterization of
\NMSOe.

\begin{lem}\label{decomp}
$\NGSMf \subseteq \MREL \cmp \DGSM = \NMSOe$.
\end{lem}

\begin{proof}
Let $\cM$ be a 2gsm, finite visit for constant $k$;
each pair $(w,z)$ in the transduction realized by $\cM$
can be computed by a $k$-visiting computation.

We may decompose the behaviour of $\cM$ on input $w$ as follows.
First, a relabelling of $\tape{w}$ guesses a string of
$k$-visiting sequences, one for each position of the input tape.
Then, a 2dgsm verifies in a left to right scan whether the
string specifies a valid computation, a track,
of $\cM$ for $w$,
cf. Proposition~\ref{regular-visiting}.
If this is the case, the 2dgsm returns to the left tape marker
$\ltape$ and simulates $\cM$ on this input, following
the $k$-visiting computation previously guessed.

When changing from one tape position to a neighbouring
position, the 2dgsm records the `crossing number' of that move,
i.e., the number of times it crossed the border
between these two tape positions (in one direction
or another). The crossing number can be read by inspecting the
directions of the moves stored in the visiting sequence.
It is used to `enter' the next visiting sequence at the right
visit, cf. Figure~\ref{fig-vis2}.
\end{proof}

\begin{thm}\label{NMSOe}
$\NMSOe
= \NGSMf \cmp \NGSMf$.
\end{thm}

\begin{proof}
By the last	lemma,
$\NGSMf \subseteq \NMSOe$.
As the right-hand side of this inclusion
 is closed under composition (Proposition~\ref{grNMSO-closed})
 we have the inclusion
$\NGSMf\cmp\NGSMf \subseteq \NMSOe$.

According to
 Theorem~\ref{NMSO=REL;DGSM},
$\NMSOe$ equals $\MREL \cmp \DGSM$.
The inclusion from left to right follows
from the fact that both
$\MREL	\subseteq \NGSMf$
and
$\DGSM \subseteq \NGSMf$.
\end{proof}

It is instructive to note that this characterization implies
the (apparently new) result that
$\NGSMf \cmp \NGSMf$ is closed under composition.
This should be contrasted to the fact that $\NGSMf$ itself is not
closed under composition. This follows from the observation
from the preceding section,
that the relation $m$ from Example~\ref{not-gsm}
does not belong to $\NGSM \supseteq \NGSMf$ (Lemma~\ref{lem:not-gsm}).
As we have observed,
it can be realized as combination
of two 2gsm's, the first one nondeterministically
changing a string
$a^n$ to a string $w\in \{a,b\}^*$ with $|w|=n$,
the second one duplicating $w$ into $w\# w$.
Both of these 2gsm's are finite visit.
(Alternatively, by Example~\ref{not-gsm},
$m\in \NMSOe$ which equals $\NGSMftwee$ as we just have seen.)

The families $\DGSM$, $\NGSMf$, and $\NGSMftwee$
form a hierarchy of transductions.
However, as far as their output languages are concerned
(ranges, or equivalently, with regular input)
these three families are equally powerful \cite{kiel,gre78b}.

\medskip
Recall that the families \NGSM\ and \NMSOe\ are incomparable,
see the discussion preceding Lemma~\ref{lem:not-gsm}.
We have a surprising characterization for their intersection.

\begin{thm}\label{intersection}
$\NGSM \cap \NMSOe = \NGSMf$.
\end{thm}

\begin{proof}
Obviously $\NGSMf \subseteq \NGSM$, while
$\NGSMf \subseteq \NMSOe$ by Theorem~\ref{NMSOe},
which proves the inclusion from right to left.

The reverse implication is immediate from Lemma~\ref{finitary}:
recall that transductions in \NMSOe\ are finitary because the
number of parameter valuations is finite.
\ \ \ \ 
\end{proof}

Combining this theorem and the related Lemma~\ref{finitary},
we obtain that a 2gsm string transduction is
mso definable if and only if it is finitary.
This generalizes a similar result of Courcelle
\cite[Proposition~6.1]{msotrans}
for rational transductions
(i.e., string transductions realized by
2gsm never moving to the left).
It can be extended to arbitrary compositions of
two-way gsm's, as we shall see in our next main result,
Theorem~\ref{thm:ngsm^k}.

As a preparation to this result (and its proof)
we like to point out that `pumping' computations for finite
visit transductions (iterating suitable segments of tracks)
does not only result in duplication of parts of the output,
but may also rearrange neighbouring segments of the output.
We illustrate this with an example.

\begin{exple}\label{permutation}
The 2gsm $\cM$ has states $1$ to $6$, initial state $1$,
final state $6$, and transitions
$(p,\sigma,\; q,\alpha,\epsilon, \; p,\empt,0 )$
where the move $q,\alpha,\epsilon$ for each pair $p\in\{1,2,\dots,5\}$,
 $\sigma\in \{\ltape,a,b,\rtape\}$
is given in the following matrix.

\[
\begin{array}{r|*5c}
       & 1 & 2 & 3 & 4 & 5 \\\hline
\ltape & 1,\empt,+1  &	3,b,0  &  3,\empt,+1  &  5,b,0	&  5,\empt,+1 \\
   a   & 1,a,+1  &  2,a,-1  &  3,a,+1  &  4,a,-1  &  5,a,+1  \\
   b   & 2,\empt,-1  &	4,\empt,-1  &  1,\empt,+1  &  5,\empt,+1  &  3,\empt,+1  \\
\rtape & 2,c,0	&  2,\empt,-1  &  4,c,0  &   4,\empt,-1  &  6,\empt,0
\end{array}
\]

(Note that the machine is nondeterministic in our setting,
but is obtained by adding dummy alternatives to a deterministic
automaton in the 5-tuple framework, see
Section~\ref{sect:2way}.)

On each segment of $a$'s of the input
$\cM$ makes five passes in states $1$ to $5$,
each in alternate directions, while copying the
letters to the output.

On a letter $b$ the machine does not generate output,
but it performs a permutation of the order in which the two
neighbouring segments of $a$'s are read.
This is best explained by looking at the computations on the
input strings  $a^3b^ia^2$,  $i=0,1,2$	as depicted
in Figure~\ref{fig:permutation}.
The output strings for these inputs are given in the following table.

\[
\begin{array}{r@{}c@{}l|c}
\multicolumn{3}{c}{\mbox{input string}} & \mbox{output string} \\\hline
\rule{0ex}{2.5ex}a^5 = {}& a^3b^0a^2 &  &  a^5ca^5ba^5ca^5ba^5 = 
                        a^3 (a^2ca^2) (a^3ba^3) (a^2ca^2) (a^3ba^3) a^2 \\
 &   a^3b^1a^2 &      & a^6ba^5ca^5ba^5ca^4 =
                        a^3 (a^3ba^3) (a^2ca^2) (a^3ba^3) (a^2ca^2) a^2 \\
 & a^3b^ia^2 &\mbox{, } i\ge 2 & a^6ba^6ba^5ca^4ca^4 =
                        a^3 (a^3ba^3) (a^3ba^3) (a^2ca^2) (a^2ca^2) a^2
\\\hline
\end{array}
\]


\begin{figure}[p]
\centerline{\unitlength 0.50mm
\begin{picture}(20,140)(0,-20)
\input 2n-zigl.gtp
\end{picture}%
\begin{picture}(20,140)(0,-20)
\input 2n-ziga.gtp
\end{picture}%
\begin{picture}(20,140)(0,-20)
\input 2n-ziga.gtp
\end{picture}%
\begin{picture}(20,140)(0,-20)
\input 2n-ziga.gtp
\end{picture}%
\begin{picture}(20,140)(0,-20)
\input 2n-ziga.gtp
\end{picture}%
\begin{picture}(20,140)(0,-20)
\input 2n-ziga.gtp
\end{picture}%
\begin{picture}(20,140)(0,-20)
\input 2n-zigr.gtp
\end{picture}%
\
\begin{picture}(20,140)(0,-20)
\input 2n-zigl.gtp
\end{picture}%
\begin{picture}(20,140)(0,-20)
\input 2n-ziga.gtp
\end{picture}%
\begin{picture}(20,140)(0,-20)
\input 2n-ziga.gtp
\end{picture}%
\begin{picture}(20,140)(0,-20)
\input 2n-zigab.gtp
\end{picture}%
\begin{picture}(20,140)(0,-20)
\input 2n-zigb.gtp
\end{picture}%
\begin{picture}(20,140)(0,-20)
\input 2n-zigba.gtp
\end{picture}%
\begin{picture}(20,140)(0,-20)
\input 2n-ziga.gtp
\end{picture}%
\begin{picture}(20,140)(0,-20)
\input 2n-zigr.gtp
\end{picture}%
}

\centerline{\unitlength 0.50mm
\begin{picture}(20,140)(0,-20)
\input 2n-zigl.gtp
\end{picture}%
\begin{picture}(20,140)(0,-20)
\input 2n-ziga.gtp
\end{picture}%
\begin{picture}(20,140)(0,-20)
\input 2n-ziga.gtp
\end{picture}%
\begin{picture}(20,140)(0,-20)
\input 2n-zigab.gtp
\end{picture}%
\begin{picture}(20,140)(0,-20)
\input 2n-zigb1.gtp
\end{picture}%
\begin{picture}(20,140)(0,-20)
\input 2n-zigb2.gtp
\end{picture}%
\begin{picture}(20,140)(0,-20)
\input 2n-zigba.gtp
\end{picture}%
\begin{picture}(20,140)(0,-20)
\input 2n-ziga.gtp
\end{picture}%
\begin{picture}(20,140)(0,-20)
\input 2n-zigr.gtp
\end{picture}%
\
\begin{picture}(20,140)(0,-20)
\input 2n-zigxl.gtp
\end{picture}%
\begin{picture}(20,140)(0,-20)
\input 2n-zigab.gtp
\end{picture}%
\begin{picture}(20,140)(0,-20)
\input 2n-zigb1.gtp
\end{picture}%
\begin{picture}(20,140)(0,-20)
\input 2n-zigbm.gtp
\end{picture}%
\begin{picture}(20,140)(0,-20)
\input 2n-zigb2.gtp
\end{picture}%
\begin{picture}(20,140)(0,-20)
\input 2n-zigba.gtp
\end{picture}%
\begin{picture}(20,140)(0,-20)
\input 2n-zigxr.gtp
\end{picture}%
}
\caption{Computations for Example~\ref{permutation}}\label{fig:permutation}
\end{figure}
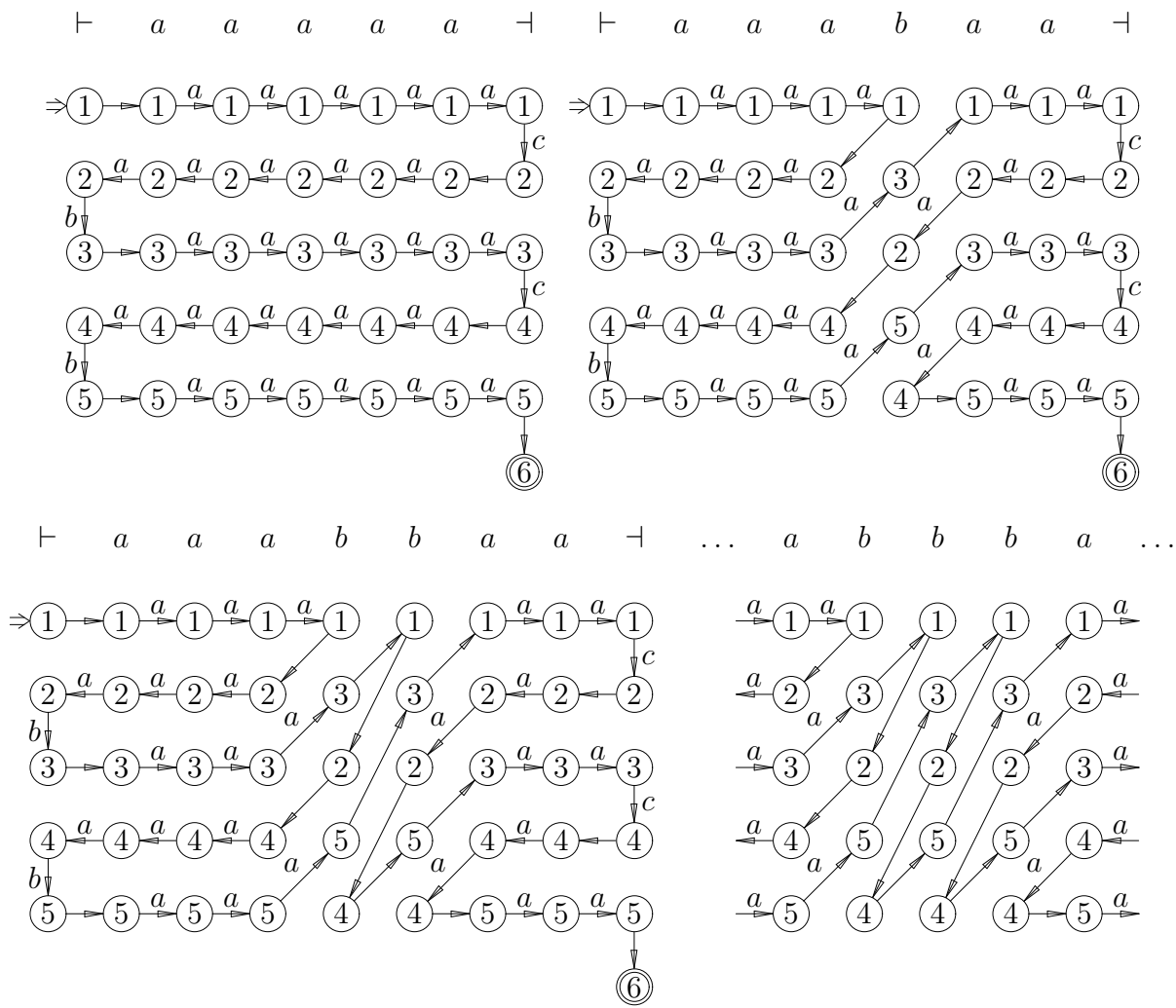

\medskip
As we have seen,
the introduction of the symbol $b$ in the input does not generate
new output. Instead, it rearranges the parts of the computation that extend
to both sides of the symbol.

Consider the boundary between two tape positions,
where we want to insert a symbol $b$.
Let $x_1,z_1, x_2, z_2,x_3,z_3$ be the strings written to the output during the
consecutive parts of the computation that visit
the left ($x_i$) and right ($z_i$) segments of the tape,
see Figure~\ref{fig:knex}.
The output generated is thus
$x_1z_1x_2z_2x_3z_3$.

Now, we introduce $b$ at the selected boundary,
and obtain the new output
$x_1x_2z_1x_3z_2z_3$.
This rearrangement of the output
can be formalized by the application of the substitution
$\sigma_b: [z_1,z_2,z_3 \leftarrow \empt,z_1,z_2z_3]$ --
where $z_i$ is a formal parameter rather than
a specific string.

The effect of introducing $bb$ can be computed by the
composition
$\sigma_b \sigma_b : [z_1,z_2,z_3 \leftarrow \empt,\empt,z_1z_2z_3]$,
which defines the rearrangement
$x_1x_2x_3z_1z_2z_3$
of the output.
Note that $\sigma_b^i = \sigma_b^2$ for $i\ge2$.
\end{exple}

\begin{figure}[p]
\centerline{
\unitlength 0.50mm
\begin{picture}(45,130)(-40,0)
\input 2n-knll.tex
\end{picture}%
\begin{picture}(10,130)(-5,0)
\input 2n-knex.gtp
\drawline(0,110)(0,0)
\end{picture}%
\begin{picture}(40,130)(-5,0)
\input 2n-knrr.tex
\end{picture}%
\
\begin{picture}(45,130)(-40,0)
\input 2n-knll.tex
\end{picture}%
\begin{picture}(10,130)(-5,0)
\input 2n-knex.gtp
\end{picture}%
\begin{picture}(20,130)
\input 2n-kn1b.tex
\end{picture}%
\begin{picture}(10,130)(-5,0)
\input 2n-knex.gtp
\end{picture}%
\begin{picture}(40,130)(-5,0)
\input 2n-knrr.tex
\end{picture}%
}

\centerline{\unitlength 0.50mm
\begin{picture}(25,130)(-20,0)
\input 2n-knlx.tex
\end{picture}%
\begin{picture}(10,130)(-5,0)
\input 2n-knex.gtp
\end{picture}%
\begin{picture}(20,130)
\input 2n-kn1b.tex
\end{picture}%
\begin{picture}(10,130)(-5,0)
\input 2n-knex.gtp
\end{picture}%
\begin{picture}(20,130)
\input 2n-kn1b.tex
\end{picture}%
\begin{picture}(10,130)(-5,0)
\input 2n-knex.gtp
\end{picture}%
\begin{picture}(20,130)(-5,0)
\input 2n-knrx.tex
\end{picture}%
\
\begin{picture}(25,130)(-20,0)
\input 2n-knlx.tex
\end{picture}%
\begin{picture}(10,130)(-5,0)
\input 2n-knex.gtp
\end{picture}%
\begin{picture}(20,130)
\input 2n-kn2b.tex
\end{picture}%
\begin{picture}(10,130)(-5,0)
\input 2n-knex.gtp
\end{picture}%
\begin{picture}(20,130)(-5,0)
\input 2n-knrx.tex
\end{picture}%
\
\begin{picture}(25,130)(-20,0)
\input 2n-knlx.tex
\end{picture}%
\begin{picture}(10,130)(-5,0)
\input 2n-knex.gtp
\end{picture}%
\begin{picture}(20,130)
\input 2n-kn2b.tex
\end{picture}%
\begin{picture}(10,130)(-5,0)
\input 2n-knex.gtp
\end{picture}%
\begin{picture}(20,130)
\input 2n-kn1b.tex
\end{picture}%
\begin{picture}(10,130)(-5,0)
\input 2n-knex.gtp
\end{picture}%
\begin{picture}(20,130)(-5,0)
\input 2n-knrx.tex
\end{picture}%
}
\caption{Visualization of rearrangements}\label{fig:knex}
\end{figure}
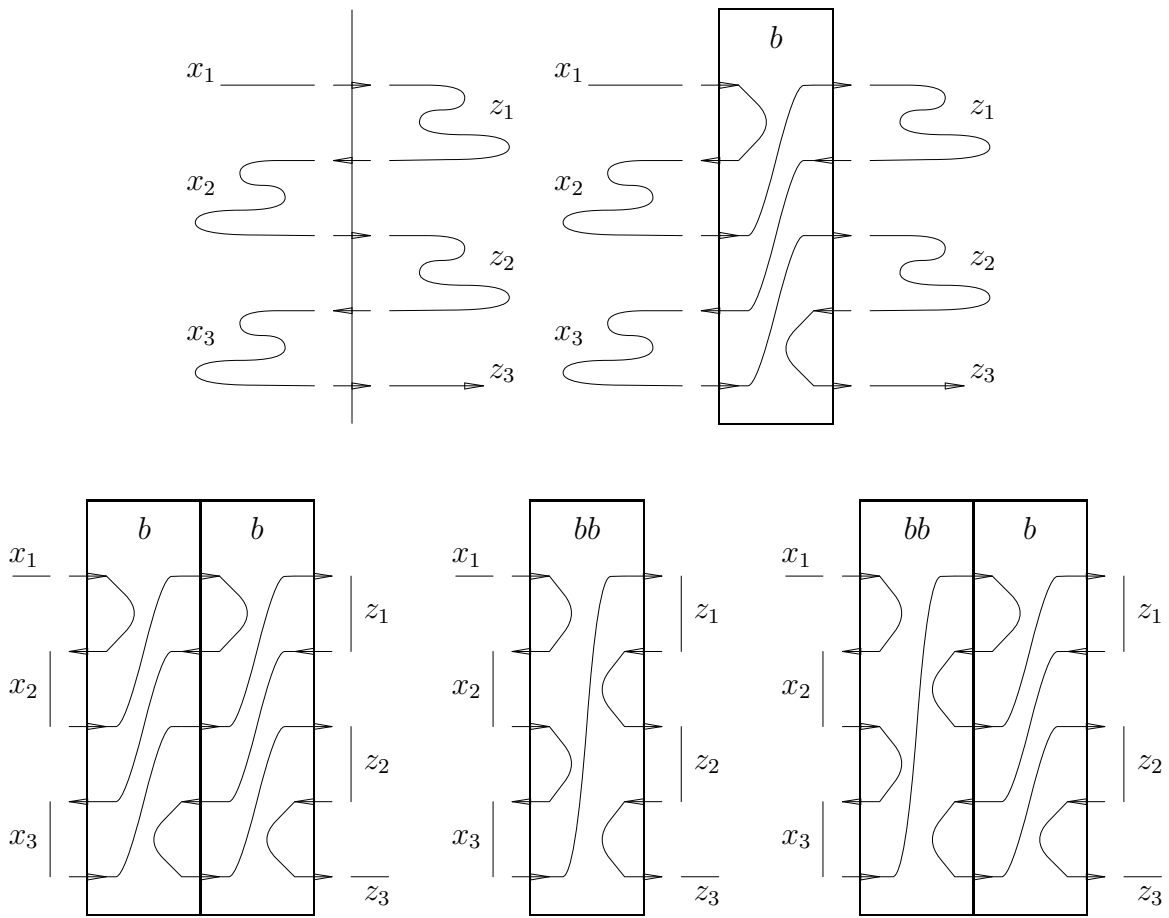

\begin{lem}
Let $m$ be a finitary string transduction,
and let $X$ be a family of string transductions.
\\
If $m \in X \cmp \NGSM \cmp \DGSM$,
then
$m \in X \cmp \MREL \cmp \DGSM$.
\end{lem}

\begin{proof}
Assume that the finitary transduction $m$ is a composition
$m = m_0 \cmp m_1 \cmp m_2$ as in the statement of the lemma;
$m_0 \in X$,
$m_1$ realized by the 2gsm $\cM_1$, and
$m_2$ realized by the 2dgsm $\cM_2$.
As to be expected, the unknown family $X$
will not feature in our arguments,
but later will enable us to apply the result in a context.
In fact, we show how to replace $m_1 \cmp m_2$
by $\breve m_1 \cmp \breve m_2 \in \MREL \cmp \DGSM$
such that $m_0 \cmp m_1 \cmp m_2 = m_0 \cmp \breve m_1 \cmp \breve m_2$.
Hence $m_1 \cmp m_2$ equals $\breve m_1 \cmp \breve m_2$ 
on the range of $m_0$.

\smallskip
Reconsider the proof of Lemma~\ref{decomp},
where a $k$-visit 2gsm is decomposed into a relabelling
that guesses a
$k$-visiting sequence for each position of the input tape,
and
a 2dgsm that verifies in a single left-to-right pass 
whether the resulting string
defines a $k$-track, and then
deterministically simulates the specified computation for the
original input.
Alternatively, by combining the verification phase
with the relabelling,
we may decompose the $k$-visit 2gsm
into a one-way gsm that nondeterministically writes a
$k$-track, and a 2dgsm simulating the computation.

We apply that new decomposition to $\cM_2$,
and immediately observe that the first phase
(guessing and writing a track) can be performed by
$\cM_1$ using a straightforward direct product construction.

Summarizing: we have replaced the composition $m_1 \cmp m_2$
by a new composition $m'_1 \cmp m'_2$
realized by $\cM'_1$ followed by $\cM'_2$,
where $\cM'_1$ is a 2gsm that writes valid tracks
for the 2dgsm $\cM'_2$.
Let $\cM'_2$ be $k$-visit.

\smallskip
We continue by demonstrating that we need not consider all
computations of $\cM'_1$, instead it suffices to put a bound on the
number of visits that the machine makes to each of the positions
of its input. This will change the transduction $m'_1$ realized by
$\cM'_1$, but not the composition $m_0 \cmp m'_1 \cmp m'_2$
(due to $m$ being finitary).

Consider the behaviour of $\cM'_1$ on input $w$,
where $w$ is in the range of $m_0$.
Fix a position on the tape $\tape{w}$ and a state of $\cM'_1$,
and split
the output of $\cM'_1$ during the computation into
segments, corresponding to the consecutive visits
to the selected position in the selected state.
$\cM'_1$ writes
$x y_1 y_2 \cdots y_t z$
where $y_i$ is written during the excursions in between
consecutive visits.
We assume $t\ge 1$.

Returning to the same position and state,
each of the excursions
can be repeated in (or omitted from) the computation
of $\cM'_1$, so the machine may
produce every string
$x y z$, $y \in \{ y_1, y_2, \dots, y_t \}^*$
as possible output on input $w$.
By our previous construction, each output of $\cM'_1$
  forms a $k$-track
for the second machine $\cM'_2$.
This implies
 that $\cM'_2$ does not generate output during
any of its visits  to the segments $y_i$,
as $m$
is supposed to be finitary.

At first glance, the excursion of $\cM'_1$ writing $y=y_1 \cdots y_t$
can be omitted:
the second machine $\cM'_2$ does not generate output
when it visits the segment $y$ during its simulation of the
specified computation.
However,
the previous example shows that $y$
(or in fact any segment $y_i$)
may have its effect on the
output of $\cM'_2$ by rearranging  parts of the adjacent computation that
leave the segment $y$ (to the left or to the right)
in order to return there later.

We consider the computation of $\cM'_2$ specified by the track $xyz$
from the viewpoint of the segment $y$.
Starting from the leftmost position of $x$, the computation enters
$y$ from the left.
Before leaving the segment for the last time,
the computation makes several tours outside $y$.

Such a tour of $\cM'_2$ to the left of the segment $y$,
in $x$,
corresponds to two consecutive visits $(\mineps, p, -1, \empt)$
and $(+1, p', \pluseps', \empt )$
in the first visiting sequence of $y$,
meaning the computation leaves the segment to the left in state $p$,
returning there later in state $p'$.
A  symmetric  observation holds for tours to the right, in $z$, 
and consecutive visits in the last visiting sequence of $y$.

Hence,
the relative order of those tours that leave to the left is fixed by the
last visiting sequence of $x$, similarly for the tours to
the right.
The relative order of all tours (left and right taken together)
is determined by the segment $y$. Replacing $y$ by another
string in
$\{ y_1, y_2, \dots, y_t \}^*$
will not change the tours in $x$ and $z$,
but it may rearrange the relative order of tours to the
left and tours to the right.

A visiting sequence for $\cM'_2$ contains at most $k$ visits.
Hence, there are less than $k$
tours to each side of the segment. Together these at most $2k$ tours
may be ordered 
 in less than $\kappa = {2k  \choose k}$ ways
(the orders of the tours at the same side of the segment are fixed).

Now we are able to apply a pumping argument to the segment
$y=y_1 \cdots y_t$.
If $t > \kappa$, then
two of the prefixes $y_1 \cdots y_{i_1}$, $y_1 \cdots y_{i_2}$,
$i_1 < i_2$,
define the
same rearrangement on the adjacent tours,
and thus we may replace $y_1 \cdots y_{i_2}$ 
by $y_1 \cdots y_{i_1}$  in
the output $xyz$ of $\cM'_1$.
The resulting track $xy_1 \cdots y_{i_1}y_{i_2+1} \cdots y_t z$  
defines a computation for $\cM'_2$ that
results in the same output as the original track $xyz$.
Thus, we may assume that $t \le \kappa$.

Consequently, we allow for all possible  rearrangements,
and hence for all possible outputs of $\cM'_2$,
 by
taking $\kappa$ as the bound on the
number of visits of $\cM'_1$ to a fixed position
in a fixed state.

Now that we have limited the number of visits of $\cM'_1$
to $\kappa$ times the size of its state set,
we can replace $\cM'_1$ by a decomposition in $\MREL\cmp \DGSM$,
using again the argumentation of Lemma~\ref{decomp}.
Thus, 
 $m_1' \cmp m'_2$ is replaced by a composition
in $(\MREL \cmp \DGSM) \cmp \DGSM$.
The result  follows, as $\DGSM$ is closed under composition,
Proposition~\ref{prop:2dgsm-composition}.
\end{proof}

The variable family $X$ in the previous result allows us to
apply the lemma in the context of an arbitrary sequence of
2gsm transductions.

\begin{thm}
Let $m$ be a string transduction, and let $k\ge 1$.
\\
If $m \in \NGSM^k$, and $m$ is finitary,
then
$m \in	\MREL\cmp\DGSM$.
\end{thm}

\begin{proof}
Observe that
$\NGSM \cmp \MREL \subseteq \NGSM$
by an obvious construction.

Let $k\ge 1$.
Assume that $m \in \NGSM^k \cmp \DGSM$ is finitary.
We have by the previous lemma,
$m \in \NGSM^{k-1} \cmp \MREL \cmp \DGSM$,
which equals $\NGSM^{k-1} \cmp \DGSM$ for $k> 1$
(and which equals $\MREL \cmp \DGSM$ for $k=1$).

Hence, by induction on $k$,
 $m \in \NGSM^k \cmp \DGSM$ implies $m \in \MREL \cmp \DGSM$,
 for a finitary string transduction $m$.
As $\NGSM^k \cmp \DGSM \supseteq \NGSM^k$,
the theorem follows.
\end{proof}

\begin{thm}\label{thm:ngsm^k} Let $m$ be a string transduction.
Then
\\
$m\in \NMSOe$ iff
$m\in \bigcup_{k\ge1}\NGSM^k$ and $m$ is finitary.
\end{thm}

\begin{proof}
By Theorem~\ref{NMSOe}, $\NMSOe = \NGSMftwee \subseteq \NGSM^2$.
Additionally, elements of \NMSOe\ are necessarily finitary.
This proves the implication from left to right.
The reverse implication follows from the last result and the
characterization  $\NMSOe= \MREL\cmp \DGSM$ from Theorem~\ref{NMSO=REL;DGSM}.
\end{proof}

It is shown in \cite[Theorem~4.9]{eng82} that every functional
transduction in $\bigcup_{k\ge1}\NGSM^k$ is in $\DGSM$.
Together with Theorem~\ref{thm-logic&machine} ($\MSOe = \DGSM$)
this gives the following counterpart of Theorem~\ref{thm:ngsm^k}.

\begin{thm}\label{thm:ngsm^k-twee} Let $m$ be a string transduction.
Then
\\
$m\in \MSOe$ iff
$m\in \bigcup_{k\ge1}\NGSM^k$ and $m$ is functional.
\end{thm}

\begin{figure}
{\unitlength 1mm
\newcommand{\ver}{\rule{0ex}{2.5ex}}
\newcommand{\hor}{\rule{1ex}{0ex}}
\begin{picture}(100,80)(0,5)
\put(10,20){\framebox(90,60)[tl]{\hor\ver$\bigcup_{k\ge1}\NGSM^k$}}
\put(20,20){\framebox(80,40)[tl]{\hor\ver$\NGSM$}}
\put(40,20){\framebox(60,50)[tl]{\hor\ver$\NMSOe=\NGSMftwee$}}
\put(40,20){\makebox(60,40)[tl]{\hor\ver$\NGSMf$}}
\put(60,20){\framebox(40,26.666)[tl]{\hor\ver$\MSOe=\DGSM$}}

\put(60,10){\dashbox(70,36.666)[tr]{functional\ver\hor}}
\put(40,10){\dashbox(90,60)[tr]{finitary\ver\hor}}
\end{picture}}
\caption{Relationships between our main families of
transductions}\label{fig:Venn}
\end{figure}
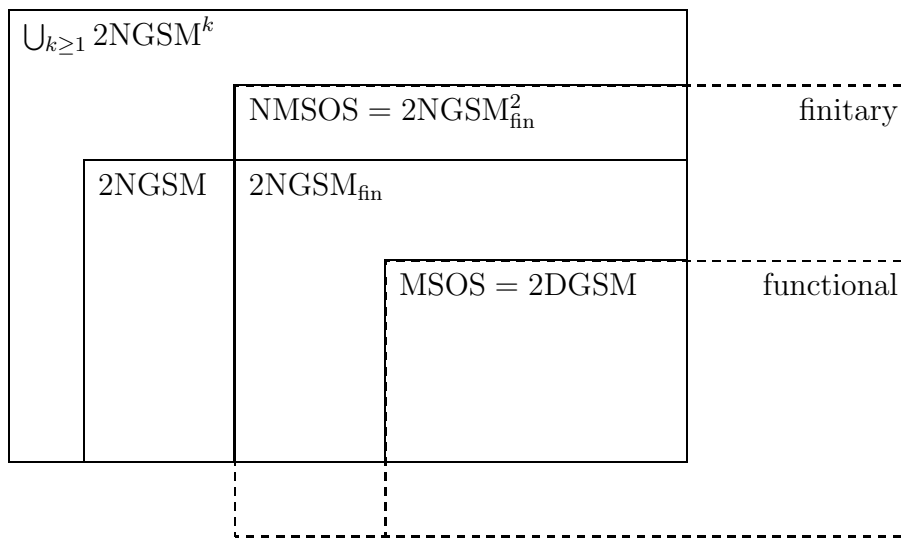

A Venn diagram is given in Figure~\ref{fig:Venn}, page~\pageref{fig:Venn}.
It illustrates the results from
Lemma~\ref{finitary},
and Theorems~\ref{intersection}, \ref{thm:ngsm^k},
 and \ref{thm:ngsm^k-twee}.

\subsection{Hennie machines}

Extending a finite visit 2gsm with the possibility
to rewrite the contents of the cell of the input tape
that it is visiting,
we obtain the {\em Hennie machine},
introduced in \cite{hennie} as an accepting device,
and considered as transducer in \cite{rajlich}
(under the name `bounded crossing transducer').
Alternatively,
a Hennie machine is a linear bounded automaton
(as transducer, so equipped with a one-way output tape)
that is finite visit.
We find it, somewhat disguised, in \cite{gre78b}
as `one way finite visit preset Turing machine',
where the `preset working tape' should be interpreted as input tape,
and the `one way input tape' as output tape.

It should be clear how to extend our basic 2sm model to allow
for writing on the input tape, thus we will refrain from giving
the full   10-tuple formalization.
The families of string transductions realized by
nondeterministic and deterministic Hennie machines
are denoted by \NHM\ and \DHM, respectively.

\begin{exple}
Once again consider our running nondeterministic example
(cf. Example~\ref{not-gsm})
\[
m = \{\; (a^n, w\# w) \mid n\ge 0, w\in\{a,b\}^*, |w|=n \;\}.
\]
It can be realized by a Hennie machine moving in two
consecutive left-to-right passes over the input.
First it nondeterministically rewrites the input $a^n$
into a string $w$ with $|w|=n$, while writing this string to
the output tape,
then it writes $w$ again to the output, copying it from the
rewritten input tape.
Obviously, the machine is 3-visit.
\end{exple}

\begin{thm}\label{thm:NHM}
$ \NMSOe = \NHM$.
\end{thm}

\begin{proof}
In view of
 Theorem~\ref{NMSO=REL;DGSM}
it suffices to prove the equality
$\NHM = \MREL\cmp\DGSM$.

The inclusion of
$\NHM$ in $\MREL \cmp \DGSM$
can be proved as Lemma~\ref{decomp},
which states this inclusion for $\NGSMf$:
the relabelling guesses a string of visiting sequences for the
computation of the Hennie machine on the input string;
the 2dgsm verifies that this string is a track and simulates the computation.
Note that a visiting sequence of a Hennie machine should also record
the symbol at the position of the input tape at each visit.
It is straigthforward to adapt the notions of visiting sequence 
and $k$-track in this way,
such that Proposition~\ref{regular-visiting} still holds
(see \cite{gre78a,gre78b,birget}).

The reverse inclusion is almost immediate.
In two phases the Hennie machine may simulate the composition,
first writing the image of the marked relabelling on the tape,
and then simulating the 2dgsm on this new tape.
There is a minor technicality: for a given input $w$
the initial tape contains
$\tape{w}$, and the Hennie machine is supposed to overwrite this
string with its relabelling and add two new tape markers
(for the simulation of the 2dgsm).
Instead, it keeps the relabelling of the tape markers in its
finite state memory, rather than overwriting them.
\end{proof}

Restating the above result as
$\NHM = \MREL\cmp\DGSM$,
it generalizes the result of Rajlich
\cite[Theorem~2.1]{rajlich}
that the output languages of nondeterministic Hennie machines
equal
the output languages of two-way
deterministic generalized sequential machines,
see also \cite[Thm~2.15(2)]{gre78a}.

The above demonstration of the inclusion
$\MREL\cmp\DGSM \subseteq \NHM$
can easily  be extended to  a proof of
$\NHM \cmp \NHM \subseteq \NHM$.
A Hennie machine can simulate the composition of two of its colleagues
by writing the visiting sequences of the first machine onto the
input tape. The output tape is contained in this string,
conveniently folded over the input tape,
ready to be used by the second machine.

We have, however, the closure  of \NHM\ under composition for free
as a consequence of the above characterization
and  Proposition~\ref{grNMSO-closed}.

\begin{cor}
\NHM\ is closed under composition.
\end{cor}

In \cite{chytil} it is noted that the inclusion
$\DHM \cmp \DGSM \subseteq \DGSM$
can be proved analogously to their result that
$\DGSM \cmp \DGSM \subseteq \DGSM$
(i.e., \DGSM\ is closed under composition,
Proposition~\ref{prop:2dgsm-composition}).
That of course implies the equality of the
families of transductions realized by
deterministic Hennie machines and those
realised by deterministic 2gsm.
This equality is rephrased as follows.

\begin{thm}
$ \MSOe= \DHM$.
\end{thm}

\begin{proof}
In view of Theorem~\ref{thm-logic&machine} it suffices to prove the equality
$\DHM = \DGSM$.
The inclusion $\DHM \supseteq \DGSM$ is immediate.
We demonstrate the reverse inclusion,
much along the lines as sketched in \cite{chytil},
see also \cite[Theorem~4.9]{eng82}.

By Theorem~\ref{thm:NHM},
$\NHM = \MREL \cmp \DGSM$.
Hence, any Hennie transduction	$m_H$
can be decomposed into a marked relabelling $\rho$
and a deterministic 2gsm transduction $m_2$.
We will argue that for a deterministic Hennie transduction
this (nondeterministic) marked relabelling can be
realized by a deterministic 2gsm, which shows $\DHM \subseteq \DGSM$
by the closure of \DGSM\ under composition.

Let $m_H$ be a deterministic Hennie transduction,
and let $m_H = \rho \cmp m_2$
be the decomposition as above.
Let $w$ be an input string.
As $m_H$ is functional, $m_H(w) = m_2(w')$
for {\em any} marked relabelling $w' \in \rho(w)$ that belongs
to the domain of $m_2$.
As this domain $\dom(m_2)$ is a regular language \cite{2to1a,2to1b},
a 2dgsm-rla can be constructed that finds and outputs such a marked relabelling
by one pass from left to right over the input,
using its look-around to check the remainder of the input for
 a  relabelling of the present
input symbol that leads to an element of $\dom(m_2)$.
This means that the 2dgsm-rla  looks ahead to test
the suffix of the tape	for membership in the language
$\rho^{-1}( L(\cA_q) )$,
where $\cA_q$ is a (fixed) one-way deterministic finite state automaton
accepting $\dom(m_2)$ except that the initial state is changed
to $q$ which is the state where $\cA$ would be after reading the
output generated by the 2dgsm-rla on the prefix,
including the relabelling chosen for the present symbol.
\end{proof}

\paragraph{Finale.}

In this section we have obtained a rather precize characterization
of mso definable  string transductions in terms of
Hennie transductions, both in the deterministic and in the nondeterministic
case.
Intuitively an important reason for this equivalence is the
inherent boundedness of both types of transductions:
mso definable transductions have a bound on the number of copies,
whereas Hennie machines have a bound on the number of visits
to each of the tape positions.

In case of determinism these two families are equal to the family
of transductions realized by two-way generalized sequential machines,
Theorem~\ref{thm-logic&machine}.
This should be contrasted to nondeterministic transductions,
where 2gsm are unable to record choices made during the computation,
whereas Hennie machines may use their tape for this purpose.

We summarize.

\begin{thm}\leavevmode
\begin{enumerate}
\item
$ \MSOe= \DHM = \DGSM$.
\item
$ \NMSOe= \NHM = \MREL\cmp\DGSM = \NGSMftwee$.
\end{enumerate}
\end{thm}

Now that the families $\NMSOe$ and $\NGSM$ have shown to be incomparable,
unlike their deterministic counterparts,
one may look for natural variants of the families that have the
same power.
For machines we have discussed such a variant. Indeed, by extending the
model with the power of rewriting its input tape (and at the same time
demanding the finite visit property) we obtain the Hennie transductions.
We leave it as an open problem how to introduce a variant of nondeterminism
for mso definable transductions that corresponds to 2ngsm.
Additionally, we did not consider transductions realized by one-way
transducers.
Another remaining problem of interest is the power of
first-order logic  to define string transductions
(where, in Definition~\ref{def-mso}, we assume all  formulas to
be first-order, see Example~\ref{ex-mso-edge}). 
Note that even for $C = \{1\}$  
there are first-order definable string
transductions that cannot be realized  by one-way
transducers (such as transforming a string into its
reversal). The  class of first-order definable string
transductions (with respect to \ngr)  such that 
$C =\{1\}$ 
and $\phi^{1,1}_*(x,y) = \edge_*(x,y)$ is 
characterized in \cite{lautemann}  to be the class of
all transductions that can be  realized by functional
aperiodic nondeterministic one-way  sequential
machines  (where a sequential  machine is a gsm that
outputs exactly one symbol at each step).
The equivalence of aperiodic finite state automata
and first-order logic was established in
\cite{mcnaughton}.

\begin{thebibliography}{XxXx99}

\bibitem[AHU69]{ahu69}
A.V. Aho, J.E. Hopcroft, J.D. Ullman,
A general theory of translation,
{\em Mathematical Systems Theory}
3 (1969) 193--221.

\bibitem[AhUl70]{ahul70}
A.V. Aho, J.D. Ullman,
A characterization of two-way deterministic classes of languages,
{\em Journal of Computer and System Sciences}
4 (1970) 523--538.

\bibitem[Bir96]{birget}
J.-C. Birget,
Two-way automata and length-preserving homomorphisms,
{\em Mathematical Systems Theory}
29 (1996) 191--226.

\bibitem[BlEn97]{bloem}
R. Bloem, J. Engelfriet,
A comparison of tree transductions defined by monadic second
      order logic and by attribute grammars.
      Leiden University Technical Report, 97-03, August 1997.
{\tt http://www.wi.leidenuniv.nl/TechRep/1997/tr97-03.html}

\bibitem[B\"uc60]{buchi}
J.R. B\"uchi,
Weak second-order arithmetic and finite automata,
{\em Zeitschrift f\"ur Mathematik, Logik und Grundlagen der Mathematik}
6 (1960) 66--92.

\bibitem[B\"uc62]{buchi62}
J.R. B\"uchi,
On a decision method in restricted second order
arithmetic, in:
Proc. Int. Congr. Logic, Methodology and Philosophy of Sciences 1960,
Stanford
University Press, Stanford, CA, 1962.

\bibitem[ChJ\'a77]{chytil}
M.P. Chytil, V. J\'akl,
Serial composition of 2-way finite-state transducers
and simple programs on strings,
in:
Automata, Languages and Programming, Fourth Colloquium
(A. Salomaa, M. Steinby, eds.),
{\em Lecture Notes in Computer Science}
vol. 52, Springer Verlag, 1977, pp.  135--147.

\bibitem[Cou91]{msoV}
B. Courcelle,
The monadic second-order logic of graphs V:
on closing the gap between definability and recognizability,
{\em Theoretical Computer Science} 80 (1991) 153--202.


\bibitem[Cou94]{msotrans}
B. Courcelle,
Monadic second-order definable graph transductions: a survey,
{\em Theoretical Computer Science}
126 (1994) 53--75.

\bibitem[Cou97]{cou97}
B. Courcelle,
The expression of graph properties and graph
transformations in monadic second-order logic,
in:
{\em Handbook of graph grammars and computing by graph transformation}
(G. Rozenberg, ed.),
vol. 1: Foundations,
World Scientific Publishing Co., 1997, pp. 313--400.

\bibitem[CoEn95]{hypergraphs}
B. Courcelle, J. Engelfriet,
     A logical characterization of the sets of hypergraphs defined by
hyperedge replacement grammars,
     {\em Mathematical Systems Theory} 28 (1995) 515-552.

\bibitem[Cho56]{chomsky}
N. Chomsky,
Three models for the description of language,
{\em IRE Transactions on Information Theory}
2 (1956) 113--124.

\bibitem[Don70]{doner}
J. Doner,
Tree acceptors and some of their applications,
{\em Journal of Computer and System Sciences} 4 (1970) 406--451.

\bibitem[Ebi95]{diekroz}
W. Ebinger,
Logical definability of trace languages,
Appendix to Chapter~10, in:
{\em The Book of Traces},
V. Diekert, G. Rozenberg (eds.), World Scientific, 1995.

\bibitem[Elg61]{elgot}
C.C. Elgot,
Decision problems of finite automata design and related arithmetics,
{\em Transactions of the American Mathematical Society}
98 (1961) 21--52.

\bibitem[Eng77]{eng77}
J. Engelfriet,
Top-down tree transducers with regular look-ahead,
{\em Mathematical Systems Theory}
10 (1977) 289--303.

\bibitem[Eng82]{eng82}
J. Engelfriet,
Three hierarchies of transducers,
{\em Mathematical Systems Theory}
15 (1982) 95--125.

\bibitem[Eng91a]{joost}
J. Engelfriet,
A characterization of context-free NCE graph languages by monadic
second-order logic on trees,
{\em Graph Grammars and Their Application to Computer Science}
(H. Ehrig, H.-J. Kreowski, G. Rozenberg, eds.),
{\em Lecture Notes in Computer Science}
vol. 532,  Springer Verlag, 1991, pp. 311--327.

\bibitem[Eng91b]{iterated}
J. Engelfriet,
Iterated stack automata and complexity classes,
{\em Information and Computation}
95 (1991) 21--75.

\bibitem[Eng97]{joost-handbook}
J. Engelfriet,
     Context-free graph grammars,
     in:  {\em Handbook of Formal Languages}
     (G. Rozenberg, A. Salomaa, eds.),
     vol.~3: Beyond Words, Springer-Verlag, 1997, pp. 125-213.

\bibitem[EnHe91]{heyker}
J. Engelfriet, L.M. Heyker,
      The string generating power of context-free hypergraph grammars.
      {\em Journal of Computer and System Sciences} 43 (1991) 328--360.
      
\bibitem[EnHo99]{icalp}
J. Engelfriet, H.J. Hoogeboom,
Two-way finite state transducers
and monadic second-order logic,
in: 26-th International Colloquium on 
Automata, Languages and Programming, 
{\em Lecture Notes in Computer Science}, vol. 1644,
Springer Verlag, 1999.

\bibitem[EnMa98]{maneth}
J. Engelfriet, S. Maneth,
Macro tree transducers, attribute grammars, and MSO definable tree translations.
Leiden University Technical Report, 98-09, August 1998.
{\tt http://www.wi.leidenuniv.nl/TechRep/1998/tr98-08.html}

\bibitem[EnOo97]{engooslog}
J. Engelfriet, V. van Oostrom,
      Logical description of context-free graph-languages,
{\em Journal of Computer and System Sciences} 55 (1997) 489-503.

\bibitem[ERS80]{ers80}
J. Engelfriet, G. Rozenberg, G. Slutzki,
Tree transducers, L systems, and two-way machines,
{\em Journal of Computer and System Sciences}
20 (1980) 150--202.

\bibitem[Fis69]{fis69}
M.J. Fischer,
Two characterizations of the context-sensitive languages,
IEEE Conference Record of 10th Annual Symposium on Switching and Automata Theory,
1969, pp. 149--156.

\bibitem[Gre78a]{gre78a}
S.A. Greibach,
 Visits, crosses, and reversals for
nondeterministic off-line machines,
{\em Information and Control}
36 (1978) 174--216.

\bibitem[Gre78b]{gre78b}
S.A. Greibach,
One way finite visit automata,
{\em Theoretical Computer Science}
6 (1978) 175--221.

\bibitem[Gre78c]{gre78c}
S.A.  Greibach,
Hierarchy theorems for two-way finite state transducers,
{\em Acta Informatica} 11 (1978) 89--101.


\bibitem[Hen65]{hennie}
F.C. Hennie,
One-tape, off-line Turing machine computations,
{\em Information and Control}
8 (1965) 553--578.

\bibitem[HoPa97]{htp}
H.J. Hoogeboom, P. ten Pas,
Monadic second-order definable text languages,
{\em Theory of Computing Systems}
30 (1997) 335--354.

\bibitem[HoUl67]{houl67}
J.E. Hopcroft, J.D. Ullman,
An approach to a unified theory of automata,
{\em The Bell System Technical Journal}
46 (1967) 1793--1829.
\\
also in:
IEEE Conference Record of 8th Annual Symposium on Switching and Automata Theory,
      Austin, Texas, 1967, pp. 140--147.

\bibitem[HoUl79]{houl-book}
J.E. Hopcroft, J.D. Ullman,
{\em Introduction to Automata Theory, Language, and Computation},
Addison--Wesley, Reading, Mass., 1979.

\bibitem[Kie75]{kiel}
D. Kiel,
Two-way a-transducers and AFL,
{\em Journal of Computer and System Sciences}
10 (1975) 88--109.

\bibitem[Kle56]{kleene}
S.C. Kleene,
Representation of events in nerve nets and finite automata,
{\em Automata Studies} (C.E. Shannon, J. McCarthy, eds.),
Annals of Mathematics Studies vol. 34,
Princeton University Press,
Princeton, N.J.,
1956, pp. 3--42.

\bibitem[Kur94]{kurshan}
R.P. Kurshan,
{\em Computer-Aided Verification of Coordinated Processes},
Princeton University Press, Princeton, N.J., 1994.

\bibitem[LMSV]{lautemann}
C. Lautemann, 
P. McKenzie, T. Schwentick, 
    H. Vollmer,
The descriptive complexity approach to LOGCFL,
in:
 16th Symposium on Theoretical Aspects of Computer
Science
(C. Meinel, S. Tison, eds.),
{\em Lecture Notes in Computer Science},
vol. 1563, Springer Verlag, 1999,
pp. 444--454.\\
Full version: Electronic 
Colloquium on Computational Complexity,
Report TR98-059.\\
{\tt ftp://ftp.eccc.uni-trier.de/pub/eccc/reports/1998/TR98-059/}

\bibitem[MCPi43]{mcculloch}
W.S. McCulloch, W. Pitts,
A logical calculus of the ideas imminent in nervous activity,
{\em Bulletin of Mathematical Biophysics}
5 (1943) 115--133.

\bibitem[MNPa71]{mcnaughton}
R. McNaughton, S. Papert,
{\em Counter-free automata}.
MIT Press, Cambridge, MA, 1971.

\bibitem[Myh57]{myhill}
J. Myhill,
Finite automata and the representation of events,
WADD TR-57-624,
Wright Patterson AFB, Ohio,
1957, pp. 112-137.

\bibitem[Ner58]{nerode}
A. Nerode,
Linear automata transformation,
{\em Proceedings of the American Mathematical Society}
 9 (1958) 541--544.

\bibitem[Nij82]{nijholt}
A. Nijholt,
The equivalence problem for LL- and LR-regular grammars,
{\em Journal of       Computer and System Sciences}
24 (1982) 149--161.

\bibitem[Pix96]{pixton}
D. Pixton,
  Regularity of splicing languages,
  {\em Discrete Applied Mathematics}
  69 (1996) 101--124.

\bibitem[Rab63]{rab63}
M.O. Rabin,
  Real-time computation,
  {\em Israel Journal of Mathematics}  1 (1963) 203--211.

\bibitem[RaSc59]{2to1a}
M.O. Rabin, D. Scott,
  Finite automata and their decision problems,
  {\em IBM Journal of Research and Development}
  3 (1959) 114--125.
\\
also in:
  {\em Sequential Machines: Selected Papers}
  (E.F. Moore, ed.),
  Addison-Wesley, Reading, MA, 1964, pp. 63--91.

\bibitem[Raj75]{rajlich}
V. Rajlich,
Bounded-crossing transducers,
{\em Information and Control}
27 (1975) 329--335.

\bibitem[See92]{seese}
D. Seese,
Interpretability and tree automata:
a simple way to solve algorithmic problems on graphs closely related to trees,
in: {\em Tree Automata and Languages}
(M. Nivat, A. Podelski, eds.),
Elsevier Science Publishers, 1992, 
pp. 83--114.

\bibitem[She59]{2to1b}
J.C. Shepherdson,
  The reduction of two-way automata to one-way automata,
  {\em IBM Journal of Research and Development}
  3  (1959) 198--200.
\\
also in:
  {\em Sequential Machines: Selected Papers}
  (E.F. Moore, ed.),
  Addison-Wesley, Reading, MA, 1964, pp. 92--97.

\bibitem[ThWr68]{thwr}
J.W. Thatcher, J.B. Wright,
Generalized finite automata theory with an application to a
decision problem of second-order logic,
{\em Mathematical Systems Theory} 2 (1968) 57--82.

\bibitem[Tho97]{handbook-thomas}
W. Thomas,
Languages, automata, and logic,
in:
{\em Handbook of Formal Languages}
(G. Rozenberg, A. Salomaa, eds.),
vol.~3: Beyond Words,
Springer Verlag, 1997, pp. 389--455.

\bibitem[Yu97]{yu}
S. Yu, Regular languages,
in:
{\em Handbook of Formal Languages}
(G. Rozenberg, A. Salomaa, eds.),
vol.~1: Word, Language, Grammar,
Springer Verlag, 1997, pp. 41--110.

\end{thebibliography}
\end{document}